\documentclass[a4paper,11pt]{article}
\pdfoutput=1 
\usepackage{jheppub}
\usepackage[T1]{fontenc} 
\usepackage{color}
\usepackage{fancybox}
\usepackage{amsmath}
\usepackage{amsfonts}
\usepackage{slashed}
\usepackage{amssymb}
\usepackage{indentfirst}
\usepackage{subcaption}
\usepackage{graphicx}
\usepackage{cleveref}
\usepackage[font={small,it}]{caption}
\usepackage[justification=centering]{caption}
\usepackage[numbers,sort&compress]{natbib}

\allowdisplaybreaks

\newcommand{\be}{\begin{equation}}
\newcommand{\bea}{\begin{eqnarray}}

\newcommand{\ee}{\end{equation}}
\newcommand{\eea}{\end{eqnarray}}
\newcommand{\Vol}{\mathrm{Vol}~}
\newcommand{\R}{\mathbb{R}}

\DeclareMathAlphabet{\mathpzc}{OT1}{pzc}{m}{it}
\newcommand{\nV}{\mathpzc{V}}
\newcommand{\nU}{\mathpzc{U}}

\newcommand{\tr}{\tilde r}

\title{\boldmath Conductivities from attractors}

\author{Johanna Erdmenger$^{\dagger +}$, Daniel Fern\'andez$^{\dagger \star}$, Prieslei Goulart$^{\dagger *}$ and Piotr Witkowski$^{\dagger}$}

\affiliation{$^{\dagger}$Max-Planck-Institut f\"{u}r Physik (Werner-Heisenberg-Institut)\\
F\"{o}hringer Ring 6, D-80805 Munich, Germany}
\affiliation{$^+$ Institut f\"ur Theoretische Physik und Astrophysik, Julius-Maximilians-Universit\"at W\"urzburg, \\ Am Hubland, 97074 W\"urzburg, Germany} 
\affiliation{$^{\star}$University of Iceland, Science Institute, Dunhaga 3, 107 Reykjav\'ik, Iceland}
\affiliation{$^{*}$Instituto de F\'{i}sica Te\'{o}rica, UNESP-Universidade Estadual Paulista\\ R. Dr. Bento T. Ferraz 271, Bl. II, S\~{a}o Paulo 01140-070, SP, Brazil}

\emailAdd{jke@mppmu.mpg.de}
\emailAdd{danielf@mppmu.mpg.de}
\emailAdd{prieslei@ift.unesp.br}
\emailAdd{piotrw@mppmu.mpg.de}

\abstract{In the context of applications of the AdS/CFT correspondence to condensed matter physics, we compute conductivities for field theory duals of dyonic planar black holes in 3+1-dimensional Einstein-Maxwell-dilaton theories at zero temperature. We combine the near-horizon data obtained via Sen's entropy function formalism with known expressions for conductivities. In this way we express the conductivities in terms of the extremal black hole charges. We apply our approach to three different examples for dilaton theories for which the background geometry is not known explicitly. For a constant scalar potential,  the  thermoelectric conductivity explicitly scales as $\alpha_{xy}\sim N^{3/2}$, as expected. For the same model, our approach yields a finite result for the heat conductivity $\kappa/T \propto N^{3/2}$  even for $T \rightarrow 0$. 

}


\begin{document} 
\hfill{MPP-2016-292}
\maketitle

\section{Introduction}

Holography provides a simple way of computing conductivities in models for condensed matter systems. Applying electric fields and thermal gradients induces linear perturbations about the black hole. The matrix of thermoelectric conductivities is obtained by solving the linearized perturbation equations. The most commonly studied case in the literature deals with electrically charged Anti-de Sitter-Reissner-Nordstr\"{o}m (AdS-RN) black holes \cite{Hartnoll:2009sz}, where the electric charge is the dual of the chemical potential of the field theory. In the presence of chemical potential and magnetic field on the CFT side, the gravity dual contains a dyonic charged black hole, i.e. a black hole with both electric and magnetic charges.  The simplest example is the dyonic AdS-RN planar black hole, for which the electric ($\sigma$), thermoelectric ($\alpha, \bar{\alpha}$) and heat conductivities ($ \bar{\kappa}$) are given by 
\be \sigma^{ij}=\frac{\rho}{B}\left(\begin{array}{cc}
0&1\\
-1&0\\
\end{array}\right) , \,\,  \alpha^{ij}=\overline{\alpha}^{ij}=\frac{s}{B}\left(\begin{array}{cc}
0&1\\
-1&0\\
\end{array}\right), \,\,  \overline{\kappa}^{ij}=\frac{s^{2}Tg_{4}^{2}}{B(\rho^{2}g_{4}^{4}+B^{2})}\left(\begin{array}{cc}
B&\rho g_{4}^{2}\\
-\rho g_{4}^{2}&B\\
\end{array}\right).  \label{condAdSRN} \ee 
Here, $\rho$ is the charge density,  $B$  the magnetic field, $g_{4}$  the constant coupling to the field strength and $s$  the entropy density. These conductivities were computed in \cite{Hartnoll:2007ai, Donos:2015bxe} and  are valid for all temperatures. 

As the full dyonic AdS-RN solution is known, the conductivities (\ref{condAdSRN}) can be written in terms of the charges of the black hole. For more involved applications to condensed matter systems, more elaborate supergravity theories are necessary. For instance, these may contain several scalar fields coupling to the field strength with corresponding potentials. In general, the full analytical gravity solutions for these theories are not known, and the explicit computation of these conductivities at finite temperature is not possible. Here, we give a prescription to compute  these conductivities explicitly at zero temperature via Sen's entropy function method, even for theories whose full dyonic black hole solution is not known. There are numerical investigations for many different backgrounds (see for instance \cite{Amoretti:2016cad,Erdmenger:2015qqa,Kim:2016jjk} and references therein), however these generically have the limitation that it is very hard to study very low or zero temperature backgrounds numerically in the presence of a chemical potential or magnetic field. On the gravity side, the zero temperature limit corresponds to taking extremal limit of black branes. Our present approach gives an analytical result for the conductivities precisely in this limit, at least for the supergravity actions considered here.

Recently, a novel holographic framework to compute DC conductivities was developed in \cite{Blake:2013bqa,Donos:2014uba,Donos:2014cya,Blake:2015ina,Donos:2015bxe,Lindgren:2015lia}. Here, the data relevant for computing the response of the system can be extracted solely from the horizon of the black hole. The expressions obtained in these references are very general: Given a theory with a known black hole solution, one can express the conductivities in terms of the horizon values of the scalar and gauge fields, and of the metric components. Holographic renormalization techniques \cite{Lindgren:2015lia} and a generalized Stokes equation \cite{Donos:2015bxe} were used to obtain the horizon data for dyonic backgrounds.  Explicit examples, in which the conductivities are written in terms of the charges of the black hole, were given for Reissner-Nordstr\"{o}m black holes and Q-lattices. However, analytical expressions for the conductivities for the simplest bosonic supergravity theory with one scalar field, i.e.~the Einstein-Maxwell-dilaton (EMD) theory, are still unknown.

In this present paper, we add a new ingredient to the issues discussed above by finding explicit analytical results for conductivities at zero temperature for dyonic black holes of EMD theory. We achieve this by applying Sen's entropy function method \cite{Sen:2005wa} to obtain the near-horizon data for this system in the presence of a scalar potential. Sen's  method consists of solving a set of algebraic equations, the {\it attractor equations}, which are obtained from the entropy function and whose solution gives the values of the fields at the horizon. The entropy function is constructed by evaluating the Lagrangian in the near-horizon region of the black hole. It is extremized by the solution to the attractor equations, which results in the entropy of the black hole. This method was originally applied for black holes with spherical horizons, but it can easily be used for black holes with planar horizons. Sen's approach is not commonly used in the holographic context. As far as we know, the only paper considering applications of the Sen's entropy function method in the AdS/CMT context is \cite{Astefanesei:2011pz}. More generally, attractors were discussed in relation to the holographic entanglement entropy in \cite{Kundu:2012jn}.

In this paper we focus on dyonic black holes, as these are examples for which obtaining zero temperature results is possible but particularly hard: No closed-form solutions are known and also numerical techniques experience problems at vanishing temperature. We note that holography of electrically charged dilatonic branes was studied in \cite{Goldstein:2009cv} for the case of constant dilaton potential and for exponential coupling to the field strength. Later, dyonic dilatonic branes were investigated in \cite{Goldstein:2010aw}. In the latter, the magnetic charge was included via $SL(2,\R)$ rotations, which was achieved by including an axion to the theory \cite{Shapere:1991ta}. More general dyonic black holes in the absence of an axion exist \cite{Kallosh:1992ii}, and they are substantially different from the ones obtained via $SL(2,\R)$ rotation. The most important difference is that they have a Reissner-Nordstr\"{o}m-like structure in the time and radial components of the metric, with two horizons. This guarantees the existence of the extremal limit, in contrast to the case of  dyonic solutions generated via a $SL(2,\R)$ rotation: These have a Schwarzschild-like structure in the time and radial components of the metric. This will be crucial in this present work,  in particular for the application of Sen's method.  There exist also numerical investigations about the holography of charged dilatonic branes at finite temperature  \cite{Cadoni:2011kv} for several couplings of the scalar with the field strength. 

The results that we obtain for the horizon data are written in terms of the charges of the black holes and the coupling constants of the theory. These explicit results allow us to analyze the behavior of the conductivities in field theory dual in terms of the rank of the gauge group $N$, of the magnetic field applied and of the charge density. The Hall conductivity is universal, and is  the same as the case of AdS-Reissner-Nordstr\"om, i.e.  it reads $\sigma_{xy}=\rho/B$. When the potential is just a cosmological constant, we show that the thermoelectric conductivity at zero temperature is given by
\begin{equation}
\alpha_{xy} = \frac{1}{3\gamma}\sqrt{\frac{\mathcal{Q}}{3B}}\,N^{3/2}\, ,
\end{equation}
with $Q\equiv g_{4}^{2}\rho$ the normalized charge density of the black hole and $\gamma$ a dimensionless constant which fixes the relation between metric and gauge field from supergravity. 
$g_4$ is the gauge theory coupling in the gravity action, which is chosen such that the Einstein-Hilbert 
and gauge kinetic terms have the same scaling with $N$.
The constant potential model is used as a prototype for more involved dilaton theories. In particular, we consider theories with exponential coupling to the field strengths and exponential scalar potentials, and a theory with quadratic dilaton expansion in both the coupling to the field strengths and the potential. We calculate the electric and thermoelectric conductivities for these theories as well. For all these theories, we also compute the ratio between the heat conductivities and the temperature under the assumption that this is a finite quantity at $T=0$. Surprisingly, $\kappa_{xx}/T=\kappa_{xy}/T$ for the constant scalar potential model. Moreover, the scaling with $N$ is  $\kappa_{xx}/T=\kappa_{xy}/T\sim N^{3/2}$, similarly to the other conductivities.

This paper is organized as follows: in section 2 we briefly review Sen's entropy function method. In section 3  we give an example of our approach by considering the well-known AdS-RN black hole solution, and show that the near-horizon data which we obtain using Sen's entropy function coincides with the one directly obtained from the black hole solution. In section 4 we  review the computation of conductivities in terms of horizon data, highlighting the result which are the starting point for ourw subsequent evaluation of the conductivities. Section 5 is devoted to the computation of the horizon data for some cases of Einstein-Maxwell-dilaton theories. Moreover, we express the conductivities in terms of the extremal black hole charges and gauge coupling of the potential. We briefly analyse the S-duality transformation properties of the attractor equations and conductivities.
In section 6 we conclude with an outlook. Appendix \ref{sec:appA} contains details about the black hole solutions to Einstein-Maxwell-dilaton theories in the spherical case. Appendix \ref{appuv} contains details about the numerical UV completions of some of the solutions considered.

\section{Entropy function}
The entropy function method was developed by Sen \cite{Sen:2005wa, Sen:2007qy}, and was first applied for the computation of the entropy and the study of the attractor mechanism \cite{Ferrara:1995ih, Ferrara:1996dd} for black holes with spherical symmetry. As discussed below, it  can be shown to be also valid for black holes with planar symmetry. The near-horizon geometry of an extremal dyonic black hole in four dimensions is $AdS_{2}\times {{S}}^{2}$ \cite{Kunduri:2007vf}, and in the planar limit this becomes $AdS_{2}\times \R^{2}$. The presence of both electric and magnetic charges assures that the space will have such a near-horizon geometry. When the black hole is only electrically (or magnetically) charged, the planar solution may have Lifshitz scaling symmetry as near-horizon geometry, and Sen's formalism may not be applied for these cases. 

For dyonic black holes, Sen's method provides us with an efficient way to compute the near-horizon data, i.e.~the near-horizon metric, the scalar and gauge fields on the horizon, and consequently the entropy density of these black holes. 
The starting point of the formalism is to consider that the near-horizon geometry of the planar black hole\footnote{Here we adapt the method to account for planar black holes. The original formalism deals with the spherical case only.} is $AdS_{2}\times \R^{2}$, whose general form is written as
\be 
ds^{2}=v\left(-r^{2}dt^{2}+\frac{dr^{2}}{r^{2}}\right)+wd\vec{x}^{2}, \label{one}  
\ee
where the constants $v$ and $w$ are the $AdS_{2}$ radius and the $\R^{2}$ radius\footnote{The volume of $\R^{2}$ is infinite, but we will only deal with finite densities in the whole paper.}, respectively. The curvature associated to the near-horizon  metric (\ref{one}) is 
\be R=-\frac{2}{v}. \label{curvature} \ee
The scalar and vector fields are constants for this geometry and are written as 
\be 
\phi_{s}=u_{s}, \,\,\, F^{(A)}_{rt}=e_{A}, \,\,\, F_{\theta\phi}^{(A)}=B_{A}, 
\label{fields}\ee
where ${e_{A}}$ and ${B_{A}}$ are related to the integrals of the magnetic and electric fluxes, which are in turn related to the electric and 
magnetic charges, respectively. The attractor mechanism states that the value of the scalars on the horizon of the extremal black hole is independent of any asymptotic condition at infinity. This value is completely determined by the electric and magnetic charges of the black hole. The function $f(u_{s}, v, w, e_{A}, p_{A})$  is defined as the Lagrangian density $\sqrt{- \det g}{\mathcal{L}}$ evaluated for the near-horizon geometry 
(\ref{one}) and integrated over the planar horizon variables \cite{Sen:2005wa, Sen:2007qy}, 
\be 
f(u_{s}, v_{i}, e_{A}, p_{A})=\int dx dy \sqrt{- \det g}{\mathcal{L}}. 
\ee
We extremize this function with respect to $u_{s}$, $v$, $w$ and $e_{A}$ by
\be 
\frac{\partial f}{\partial u_{s}}=0, \,\,\, \frac{\partial f}{\partial v}=0, \,\,\, \frac{\partial f}{\partial w}=0, \,\,\,  \frac{\partial f}{\partial e_{A}}=Q^{A},
\ee
where the first equation is the equation of motion for the scalar, and the second and third are the equations of motion for the metric. 
Next,  one defines
the entropy function
\be 
{\mathcal{E}}(\vec{u},\vec{v},\vec{e},\vec{q},\vec{p})\equiv 2\pi[e_{A}Q^{A}-f(\vec{u},\vec{v},\vec{e},\vec{p})]. \label{entrfct}
\ee
The equations that extremize the entropy function are
\be 
\frac{\partial {\mathcal{E}}}{\partial u_{s}}=0, \,\,\, \frac{\partial {\mathcal{E}}}{\partial v}=0,
 \,\,\, \frac{\partial {\mathcal{E}}}{\partial w}=0, \,\,\, \frac{\partial {\mathcal{E}}}{\partial e_{A}}=0\;, \label{attracteqs}
\ee
and are called the attractor equations. At the extremum, this new function equals the entropy of the black hole
\be 
S_{BH}={\mathcal{E}}(\vec{u},\vec{v},\vec{e},\vec{q},\vec{p}).
\ee
In the context of planar black holes the horizon has infinite area, so we consider the entropy density, since this is a finite quantity. The solutions of the equations (\ref{attracteqs}) are the near-horizon data that will be used later to compute the conductivities.

The attractor mechanism is independent of supersymmetry, and relies only on the near horizon geometry \cite{Sen:2005wa}, which is $AdS_{2}\times \R^{2}$ in our case. In the study of spherical black holes with $AdS_{2}\times S^{2}$ near-horizon geometry, the attractor mechanism is present even after the inclusion of $\alpha'$ corrections \cite{Sen:2005wa,Sen:2005iz,Sahoo:2006vz,Sahoo:2006rp,Alishahiha:2006ke,Chandrasekhar:2006kx}. It is now understood that the long throat of $AdS_{2}$ is the basis of the attractor phenomena \cite{Sen:2005wa,Astefanesei:2006dd,Kallosh:2006bt}. Since the $AdS_{2}\times \R^{2}$ near-horizon geometry is the starting point of the construction of Sen's entropy function, finding solutions to the attractor equations guarantees that the attractor mechanism exist for the theories analyzed.  

Notice that the entropy function method does not require the use of boundary terms. It was shown in \cite{Majhi:2015pra} that a similar formalism can be obtained by replacing the Einstein-Hilbert term of the action by a boundary term, and the results obtained are exactly the same as the ones obtained by using the original method. In this paper we will use the original method developed by Sen.

\section{Dyonic AdS-RN planar black hole\label{sec:AdSRN}}
As one example of application of the entropy function we will consider the dyonic AdS-RN case without scalars\footnote{We emphasize that there is no new result in this section. This is just an illustration of how one can obtain near-horizon data using the entropy function method. }. As we know the planar black hole solution for this model, we can compare the results obtained by applying Sen's method with the ones obtained by taking the near-horizon limit of this dyonic solution. We adopt the same units as in \cite{Hartnoll:2009sz}. The theory contains just a metric and a gauge field, and we consider also a constant potential. The action is written as 
\be S=\int d^{4}x\sqrt{-g}\left[\frac{1}{2\kappa_{4}^{2}}\left(R+\frac{6}{L^{2}}\right)-\frac{1}{4g_{4}^{2}}F_{\mu\nu}F^{\mu\nu}   \right]. \label{actionaAdSRN}\ee
Here, $L$ is the AdS$_{4}$ radius, and  $2\kappa_{4}^{2}=16\pi G_{N}$. The gauge coupling $g_4$ is assumed to scale as $\sim N^{-3/4}$. This ensures that the gravity and gauge kinetic terms have the same scaling with $N$. The relation between the gravity coupling $\kappa^{2}$, the AdS$_{4}$ radius, and the rank $N$ of the gauge group of the field theory  is given by \cite{Hartnoll:2007ai}
\be \frac{2L^{2}}{\kappa_{4}^{2}}=\frac{\sqrt{2}N^{3/2}}{6\pi}. \label{rank}\ee
The equations of motion admit the solution
\be ds^{2}=\frac{L^{2}}{u^{2}}\left(-f(u)dt^{2}+\frac{du^{2}}{f(u)}\right)+\frac{L^{2}}{u^{2}}(dx^{2}+dy^{2}), \label{metAdSRN}\ee
\be f(u)=1-\left(1+\frac{u_{+}^{2}\mu^{2}+u_{+}^{4}B^{2}}{\gamma^{2}}\right)\left(\frac{u}{u_{+}}\right)^{3}+\left(\frac{u_{+}^{2}\mu^{2}+u_{+}^{4}B^{2}}{\gamma^{2}}\right)\left(\frac{u}{u_{+}}\right)^{4}, \ee
\be F=\frac{\mu}{u_{+}} du\wedge dt+B dx\wedge dy. \ee
In this coordinate system the horizon of the black hole is located at $u_{+}$, the asymptotic region is achieved when $u\rightarrow \infty$. $B$ is the magnetic field and $\mu$ is the chemical potential. The temperature and the constant $\gamma^{2}$ are given by 
\be T=\frac{1}{4\pi u_{+}}\left(3-\frac{u_{+}^{2}\mu^{2}+u_{+}^{4}B^{2}}{\gamma^{2}}\right), \,\,\, \gamma^{2}=\frac{2g_{4}^{2}L^{2}}{\kappa_{4}^{2}}. \label{defgamma}\ee
Notice that, as $g_4 \sim N^{-3/4}$, $\gamma $ is independent of $N$. The black hole becomes extremal when
\be \frac{u_{+}^{2}\mu^{2}+u_{+}^{4}B^{2}}{\gamma^{2}}= 3. \label{extcond}\ee
In quantum field theory, the charge density $\rho$ is defined as the expectation value of the charge operator, $J^{t}$, which is the time component of a conserved current on flat space-time ($\partial_i J^i = 0$). The holographic dual of this current is the $U(1)$ gauge field, and so the standard holographic dictionary gives
\be \rho = \langle J^{t}\rangle =\frac{\delta S_{\text{on-shell}}}{\delta A_{t, \text{bdry}}}.\label{jt}\ee
This variation of the on-shell action with respect to the boundary value of the gauge field $A_{t}$  reads \cite{Donos:2014cya}
\begin{equation}
\rho = \left. \frac{\sqrt{-g}}{g^2_4} F^{tr}\right\vert_\text{bdry} \, ,\label{eq:Density0} 
\end{equation}
where the ''bdry'' subscript denotes that the expression should evaluated at the boundary. Substituting the RN action and gauge field, we obtain (see \cite{Hartnoll:2009sz})
\be \rho=\frac{2L^{2}}{\kappa_{4}^{2}}\frac{\mu}{u_{+}\gamma^{2}}. \ee
Notice that the above expression is independent of the radius, even without taking the limit $r~\rightarrow$ boundary. This important feature is exhibited in a broader class of models, and we shall comment on it later on.
As we intend to make a comparison between the quantities computed from the full extremal solution with the ones obtained via the entropy function, we write all of them in terms of the charge density, since this is the quantity that appears in the entropy function. So, using (\ref{extcond}), the value of $u_{+}$ for the extremal black hole, is given by the relation
\be \frac{1}{u_{+}^{2}}=\sqrt{\frac{1}{3\gamma^{2}}\left(B^{2}+g_{4}^{4}\rho^{2}\right)}. \ee
In order to see how the $AdS_{2}\times \R^{2}$ near-horizon geometry arises from the full extremal solution, we Taylor expand the function $f(u)$ around the horizon:
\be f(u)\approx f(u_{+})+(u-u_{+})f'(u_{+})+\frac{(u-u_{+})^{2}}{2}f''(u_{+}), \ee
where the primes define derivatives with respect to $u$. The first term in the expansion is zero due to the definition of the horizon of the black hole, and the second one is zero due to the fact that we consider extremal black holes. Notice that for the extremal black hole
\be f''(u_{+})=\frac{12}{u_{+}^{2}}. \ee
The metric (\ref{metAdSRN}) becomes
\be ds^{2}\approx \frac{L^{2}}{u_{+}^{2}}\left(-\frac{6}{u_{+}^{2}}(u-u_{+})^{2}dt^{2}+\frac{u_{+}^{2}}{6}\frac{du^{2}}{(u-u_{+})^{2}}\right)+\frac{L^{2}}{u_{+}^{2}}(dx^{2}+dy^{2}).  \ee
Defining new coordinates as
\be r \equiv u-u_{+}, \,\,\,  \tau \equiv \frac{6}{u_{+}^{2}} t,\ee
and the metric and gauge field will be
\be ds^{2}\approx \frac{L^{2}}{6}\left(-r^{2}d\tau^{2}+\frac{dr^{2}}{r^{2}}\right)+\frac{L^{2}}{u_{+}^{2}}(dx^{2}+dy^{2}),  \label{rnmetric}\ee
\be F=\frac{\rho g_{4}^{2} u_{+}^{2}}{6}dr\wedge d\tau+B dx\wedge dy. \label{rnfs} \ee
This just shows that the AdS-RN planar black hole has $AdS_{2}\times \R^{2}$ near-horizon geometry. 

We now show how these results are recovered within Sen's entropy function formalism.
For this purpose, we define the near-horizon metric and the gauge field as
\be ds^{2}= v\left(-r^{2}d\tau^{2}+\frac{dr^{2}}{r^{2}}\right)+w(dx^{2}+dy^{2}),  \ee
\be F=e\, dr\wedge d\tau+B dx\wedge dy. \ee
For this background and fields, the Lagrangian reads
\be \sqrt{-g}{\mathcal L}=\frac{1}{\kappa_{4}^{2}} \left(-w+\frac{3}{L^{2}}vw\right) +\frac{w}{2g_{4}^{2}v}e^{2}-\frac{v}{2g_{4}^{2}w}B^{2}  ,  \ee
and the entropy function is just
\be {\mathcal E}=2\pi [eQ-\int dx dy \sqrt{-g}{\mathcal L}]. \ee
Computing derivatives with respect to the fields we obtain the attractor equations
\be e=g_{4}^{2}\frac{v}{w}\tilde{Q}, \ee
\be \frac{w}{v^{2}}e^{2}+\frac{1}{w}B^{2}-\frac{6g_{4}^{2}}{\kappa_{4}^{2}L^{2}}w=0, \ee
\be \frac{2g_{4}^{2}}{\kappa_{4}^{2}}- \frac{e^{2}}{v}-\frac{v}{w^{2}}B^{2}-\frac{6g_{4}^{2}}{\kappa_{4}^{2}L^{2}}v=0, \ee
where $\tilde{Q}=Q/\Vol \R^{2}$. The solution to the system is written as
\be e=\frac{g_{4}^{2}\tilde{Q}}{2}\sqrt{\frac{\gamma^{2}}{3(g_{4}^{4}\tilde{Q}^{2}+B^{2})}}, \,\, v=\frac{L^{2}}{6}, \,\, w=L^{2}\sqrt{\frac{1}{3\gamma^{2}}(g_{4}^{4}\tilde{Q}^{2}+B^{2})}, \label{solutionattr3} \ee
where the definition of $\gamma^{2}$ is given in (\ref{defgamma}). The constant $\tilde{Q}$ is a parameter in Sen's entropy function method that is proportional to the charge density $\rho$. In order to obtain the same near-horizon we make the identification
\be \rho\equiv \tilde{Q} . \label{relQrho}\ee
Comparing the above results with (\ref{rnmetric}) and (\ref{rnfs}), we see that we have obtained exactly the same near-horizon metric and gauge fields via the entropy function. The extremal AdS-RN theory is a particular example of the model with quadratic expansion in the dilaton field, as we will discuss in subsection 5.3.

\section{DC conductivity from horizon data\label{sec:Cond}}

The computation of conductivities from horizon data was developed for theories exhibiting charge dissipation. This analysis began in the context of massive gravity \cite{Blake:2013bqa}, and subsequently was performed in the context of Q-lattices \cite{Donos:2014uba,Gouteraux:2014hca}, where the axionic term in the action is responsible for breaking translation symmetry. In all of these cases, the black hole solution considered is electrically charged. The generalization to include a dyonic black hole was recently developed in \cite{Blake:2015ina,Donos:2015bxe}. Since the conductivity expressions obtained there are the starting point for the analysis of this present paper, we review  them here.

The theory considered in \cite{Blake:2015ina,Donos:2015bxe} is given by\footnote{In our conventions, the potential appears in the action with a minus sign, and we write the gauge coupling $g_{4}^{2}$ explicitly.}
\be  S=\int d^{4}x \sqrt{-g}\left[\frac{1}{16\pi G_{N}}\left(R-\frac{1}{2}[(\partial \phi)^{2}+\Phi(\phi)\left((\partial\chi_{1})^{2}+(\partial\chi_{2})^{2}\right)]-V(\phi)\right)-\frac{Z(\phi)}{4g_{4}^{2}}F^{2}\right], \label{axionicaction}\ee
where $\phi$ is a real scalar field and the couplings $\Phi_{i}(\phi)$,  $Z(\phi)$ and $V(\phi)$ depend only on the real scalar. $F$ is an Abelian field strength. We retain the explicit $g_{4}^{2}$ dependence in the action to be able to track the $N$ dependence (recall that $g_4 \sim N^{-3/4}$ in our case).  At the moment the functions $
\Phi$, $Z$ and $V$ are arbitrary, but in order to guarantee that there exists an asymptotically AdS solution, the potential must satisfy the conditions
\begin{equation}
V(0) = -\frac{6}{L^2},~V'(0)=0,\label{PotentialCond} 
\end{equation}
where the prime denotes derivative with respect to the argument (i.e field value). The existence of an AdS solution ensures that this four-dimensional spacetime is dual to a 
three-dimensional conformal field theory. At this point it is  instructive to recall that
\begin{equation}
Z'(0) = 0 \label{eq:Z0}
\end{equation}
ensures that the AdS-RN  black hole with vanishing scalar  is a solution to the EOM. This describes a conformal field theory at finite chemical potential. However, \eqref{eq:Z0} not essential for the existence of a CFT dual \cite{Donos:2015bxe}. 

The two fields $\chi_1,~ \chi_2$ in \eqref{axionicaction}
are auxiliary scalars which may be switched on to introduce a spatial inhomogeneity, for example by setting
\be
\chi_1=k_1 x,	~ \chi_2=k_2 y .
 \ee 
For isotropic theories, the background takes the form
\be ds^{2}=-\nU\, dt^{2}+\nU^{-1}dr^{2}+e^{2\nV}(dx^{2}+dy^{2}), \label{metdyonic}\ee
\be A=a(r)dt-Bydx, \ee
\be k_1= k_2 \equiv k \, , \ee
where $\nU, \nV, a$ and $\phi$ are functions of $r$ only. The form of the action \eqref{axionicaction} is very general, and describes  the Q-lattice models in particular \cite{Donos:2013eha}. The Einstein-Maxwell-dilaton theory, which is the focus of this work, is obtained by taking $\Phi(\phi)=0$. 

The expressions for electric, thermoelectric and heat conductivities for dyonic black holes are respectively given by\footnote{The expressions given in \cite{Blake:2015ina} for these conductivities are obtained by setting $ 16\pi G_{N}=2\kappa_{4}^{2}\to1,\;g_{4}^{2}\to1 $.}

\begin{align}
\sigma_{xx}&=\left.\frac{e^{2\nV}k^{2}\Phi(2\kappa_{4}^{2}g_{4}^{4}\rho^{2}+2\kappa_{4}^{2}B^{2}Z^{2}+g_{4}^{2}Ze^{2\nV}k^{2}\Phi)}{4\kappa_{4}^{4}g_{4}^{4}B^{2}\rho^{2}+(2\kappa_{4}^{2}B^{2}Z+g_{4}^{2}e^{2\nV}k^{2}\Phi)^{2}}\right|_{r_+}, \label{elconddxx} \\
\sigma_{xy}&=\left.4\kappa_{4}^{2}B\rho\frac{\kappa_{4}^{2}g_{4}^{4}\rho^{2}+\kappa_{4}^{2}B^{2}Z^{2}+g_{4}^{2}Ze^{2\nV}k^{2}\Phi}{4\kappa_{4}^{4}g_{4}^{4}B^{2}\rho^{2}+(2\kappa_{4}^{2}B^{2}Z+g_{4}^{2}e^{2\nV}k^{2}\Phi)^{2}}\right|_{r_+}, \label{elconddxy}\\\alpha_{xx}&=\left.\frac{2\kappa_{4}^{2}g_{4}^{4}s\rho e^{2\nV}k^{2}\Phi}{4\kappa_{4}^{4}g_{4}^{4}B^{2}\rho^{2}+(2\kappa_{4}^{2}B^{2}Z+g_{4}^{2}
e^{2\nV}k^{2}\Phi)^{2}}\right|_{r_+}, \label{etconddxx} \\
\alpha_{xy}&=\left. 2\kappa_{4}^{2}sB\frac{2\kappa_{4}^{2}g_{4}^{4}\rho^{2}+2\kappa_{4}^{2}B^{2}Z^{2}+g_{4}^{2}Ze^{2\nV}k^{2}\Phi}{4\kappa_{4}^{4}g_{4}^{4}B^{2}\rho^{2}+(2\kappa_{4}^{2}B^{2}Z+g_{4}^{2}e^{2\nV}k^{2}\Phi)^{2}}\right|_{r_+}.\label{etconddxy}\\ 
\bar{\kappa}_{xx}&=\left.\frac{2\kappa_{4}^{4}g_{4}^{2}s^{2}T(2\kappa_{4}^{2}B^{2}Z+g_{4}^{2}e^{2\nV}k^{2}\Phi)}{4\kappa_{4}^{6}g_{4}^{4}B^{2}\rho^{2}+\kappa_{4}^{2}(2\kappa_{4}^{2}B^{2}Z+g_{4}^{2}e^{2\nV}k^{2}\Phi)^{2}}\right|_{r_+}, \label{hconddxx}\\
\bar{\kappa}_{xy}&=\left. \frac{4\kappa_{4}^{4}g_{4}^{4}s^{2}T\rho B}{4\kappa_{4}^{4}g_{4}^{4}B^{2}\rho^{2}+(2\kappa_{4}^{2}B^{2}Z+g_{4}^{2}e^{2\nV}k^{2}\Phi)^{2}}\right|_{r_+}.\label{hconddxy} 
\end{align}

In these expressions, $\rho$ is the charge density and $s$ is the entropy density, which are given by 
\be \rho=\left.\frac{Ze^{2\nV}a'}{g_{4}^{2}}\right|_{r_{+}}, \,\,\, s=\left. \frac{4\pi e^{2\nV}}{16 \pi G_{N}}\right|_{r_{+}}.  \label{eq:sigmaalpha} \ee
The charge density is now evaluated \emph{at the horizon} of the black hole, although it describes a field theory (boundary) quantity. This is possible due  to its radius independence (see \cite{Donos:2014cya}). In short, this feature follows from the fact that the gauge equation of motion has the form
\begin{equation}
\partial_\mu{} \left( \sqrt{-g} Z(\phi) F^{\mu{} t} \right) =0,
\end{equation}
 which  in translationally invariant set-ups implies the radial independence of $\sqrt{-g} Z(\phi) F^{r t} $. Note that (\ref{eq:Density0}) is modified  in presence of a dilatonic coupling, where it reads $\rho = \sqrt{-g}\frac{Z(\phi)}{g^2_4}F^{tr}$.

The conductivity expressions given above are valid for all temperatures. They become very simple when $\Phi(\phi)=0$. In the next sections we will consider precisely this case. Then,  the off-diagonal non-zero electric and thermoelectric conductivities are always given by
\be \sigma_{xy}=\frac{\rho}{B}, \,\,\, \alpha_{xy}=\frac{s}{B}. \label{eq:rhos} \ee
In the next section we will derive the entropy density for some specific models in which $\Phi(\phi)=0$.  The electric Hall conductivity, $\sigma_{xy}$  as given by \eqref{eq:sigmaalpha} is already expressed in terms of the electric and magnetic charges of the black hole. Below we will find a similar expression for $\alpha_{xy}$ in terms of the parameters of the black hole, i.e. the electric charge, the magnetic charge and the gauge coupling of the potential. This requires to calculate the entropy density in particular. In subsection \ref{sec:kT}, we compute the ratio between the heat conductivity and the temperature for all the theories analyzed, assuming that this is non-zero at zero temperature.

\section{DC conductivities at $T=0$ for Einstein-Maxwell-Dilaton theories}

Our strategy for obtaining explicit expressions for the conductivities will be following: First we construct Sen's entropy function for an EMD theory with coupling $Z(\phi)$ and potential $V(\phi)$  unspecified. Then we use these functions to derive the general form of the \emph{attractor equations} for this class of theories. Then, in the subsections, we analyze the equations for specific choices of $Z(\phi),~V(\phi)$ and use the solutions to compute conductivities.

The four-dimensional EMD theory with action
\be S=\int d^{4}x\sqrt{-g}\left[\frac{1}{2\kappa_{4}^{2}}\left(R-\frac{1}{2}\partial_{\mu}\phi\partial^{\mu}\phi-V(\phi)\right)-\frac{Z(\phi)}{4g_{4}^{2}}F_{\mu\nu}F^{\mu\nu}   \right], \label{actionEMD}\ee
involves the metric $g_{\mu\nu}$, a gauge field $A_{\mu}$ and a real scalar $\phi$, which is the dilaton. As stated before, this theory is obtained by taking $\Phi(\phi)=0$ in (\ref{axionicaction}). Using (\ref{one}), (\ref{curvature}) and (\ref{fields}), we compute the Lagrangian in the near-horizon region, i.e.
\be \sqrt{-g}{\mathcal{L}}=\frac{1}{16\pi G_{N}}\left(-2w-wvV(u_{D})\right)+\frac{Z(u_{D})}{2g_{4}^{2}}\left(\frac{w}{v}e^{2}-\frac{v}{w}B^{2}\right), \ee
where $u_{D}$ is the value of the dilaton field on the horizon of the black hole. The entropy function (\ref{entrfct}) is then
\be 
{\mathcal{E}}=2\pi[e_{A}Q^{A}-\Vol\R^{2}\sqrt{-g}{\mathcal{L}} ]. \label{entrfct1d}
\ee
The attractor equations for this system are
\be  \frac{Q}{\Vol\R^{2}}- \frac{Z(u_{D})}{g_{4}^{2}}\frac{w}{v}e=0,\label{QAd}\ee
\be \frac{Z(u_{D})}{2g_{4}^{2}}\left(\frac{w}{v^{2}}e^{2}+\frac{B^{2}}{w}\right) +\frac{w}{16\pi G_{N}}V(u_{D})=0,  \label{dervd}\ee
\be \frac{2}{16\pi G_{N}}- \frac{Z(u_{D})}{2g_{4}^{2}}\left(\frac{1}{v}e^{2}+\frac{v}{w^{2}}B^{2}\right) +\frac{v}{16\pi G_{N}}V(u_{D})=0,\label{derwd} \ee
\be -\frac{1}{2g_{4}^{2}}\frac{\partial Z(u_{D})}{\partial u_{D}}\left(\frac{w}{v}e^{2}-\frac{v}{w}B^{2}\right)+\frac{wv}{16\pi G_{N}}\frac{\partial V(u_{D})}{\partial u_{D}}=0. \label{derphid}\ee
Using (\ref{QAd}) we  eliminate $Q$ from (\ref{entrfct1d}), and obtain
\be {\mathcal{E}}=2\pi \Vol\R^{2}\left[\frac{1}{(16\pi G_{N})}\left(2w+wvV(u_{D}\right)+\frac{Z(u_{D})}{2g_{4}^{2}}\left(\frac{w}{v}e^{2}+\frac{v}{w}B^{2}\right) \right].   \ee 
We  combine equations (\ref{dervd}) and (\ref{derwd}) in such a way as to obtain
\be V(u_{D})=-\frac{1}{v},\label{sum} \ee
\be \frac{Z(u_{D})}{2g_{4}^{2}}\left(\frac{e^{2}}{v^{2}}+\frac{B^{2}}{w^{2}}\right)=\frac{1}{(16\pi G_{N})}\frac{1}{v},   \label{sub}\ee
and replacing this in (\ref{entrfct1d}), we obtain
\be 
{\mathcal{E}}=\frac{4\pi w \Vol\R^{2}}{16\pi G_{N}}=\frac{w \Vol\R^{2}}{4G_{N}}=\frac{A}{4G_{N}}. 
\ee
which shows that this is in agreement with the Hawking formula. 

We now aim at using  equations (\ref{QAd}, \ref{dervd}, \ref{derwd}, \ref{derphid}) to evaluate the conductivity formulae given in the previous section, i.e. (\ref{elconddxx}, \ref{elconddxy}, \ref{etconddxx}, \ref{etconddxy}, \ref{hconddxx}, \ref{hconddxy}). This requires 
to  establish a map between the metric elements and fields appearing in Sen's formalism with the metric and fields appearing in the conductivity formulae. One example, as will be shown below, is the the ratio $e/v$. Both the electric field $e$ and the $AdS_{2}$ radius $v$ appears in Sen's formalism. This ratio is related to the quantity $a'(r)$ appearing in the expressions for the conductivity through equation (\ref{releva}). We will clarify  these issues  in what follows. 

The field strength for the theory (\ref{axionicaction}) is written in terms of the quantities $a'(r)$ and $B$ such that
\be F=a'(r)dr\wedge dt+B dx\wedge dy.  \ee
This expression is valid anywhere in the bulk. Let us now consider 
the field strength and the metric in the near-horizon region. Notice that the function $\nU(r)$ appearing in (\ref{metdyonic}) can be Taylor expanded around the horizon of the black hole as
\be \nU(r)\approx \nU(r_{H})+(r-r_{H})\nU'(r_{H})+\frac{(r-r_{H})^{2}}{2}\nU''(r_{H})+\mathcal{O}(r^{3}).  \label{taylor}\ee
The first term in this expansion vanishes at the horizon by definition, and the linear term vanishes for extremal black holes. So in the near-horizon region the metric is written as
\be ds^{2}=-\frac{(r-r_{H})^{2}}{2}\nU''(r_{H})dt^{2}+\frac{2}{(r-r_{H})^{2}\nU''(r_{H})}dr^{2}+e^{2\nV(r_{H})}(dx^{2}+dy^{2}).   \ee
In order to see how the $AdS_{2}\times \R^{2}$ geometry emerges, we need to choose an appropriate coordinate system. For our case this is
\be r-r_{H}\rightarrow \tilde{\rho}, \,\,\, t\rightarrow \frac{2\tau}{\nU''(r_{H})}, \label{change} \ee 
so that
\be ds^{2}=\frac{2}{\nU''(r_{H})}\left(-\tilde{\rho}^{2}d\tau^{2}+\frac{d\tilde{\rho}^{2}}{\tilde{\rho}^{2}}\right)+e^{2\nV(r_{H})}(dx^{2}+dy^{2}),  \label{adsmetr} \ee
As we can see, the metric has $AdS_{2}\times \R^{2}$ as its near horizon geometry, as expected. A direct comparison with (\ref{one}) shows that the term multiplying the $AdS_{2}$ part of this metric is identified with $v$, and the term multiplying the $\R^{2}$ part is identified with $w$. Also, under the change of coordinates (\ref{change}), the field strength changes, since it transforms as a tensor. It becomes  
\be F=\frac{2a'(r_{H})}{\nU''(r_H)}d\tilde{\rho}\wedge d\tau + Bdx\wedge dy.  \ee
Again, a direct comparison with (\ref{fields}) tells us that the $(\tilde{\rho}\tau)$ component of the field strength is identified with $e$, and the angular part containing the magnetic field is the same. This provides us with the quantities that map the horizon data obtained via Sen's entropy function to the quantities appearing in the expressions (\ref{elconddxx}, \ref{elconddxy}, \ref{etconddxx}, \ref{etconddxy}, \ref{hconddxx}, \ref{hconddxy}). They are written as 
\be v=\frac{2}{\nU''(r_{H})}, \,\,\, w=e^{2\nV(r_{H})} , \,\,\, e=\frac{2a'(r_{H})}{\nU''(r_H)}=va'(r_{H}).\label{identification}\ee
 
The main ingredient used in the computation of the conductivities from section \ref{sec:Cond} is that the currents in the boundary theory are related to quantities in the bulk that do not depend on the radial coordinate, see \cite{Donos:2014cya} and \cite{Blake:2015ina} and references therein. This means that the boundary charge density for the theory (\ref{axionicaction}) is given by  the horizon expression  
\be \rho = \frac{Z( u_{D})wa'(r_{H})}{g_{4}^{2}} \, , \label{eqqrho} \ee
which replaces \eqref{eq:Density0}. In our notation $A_{t}$ is just $a$. This vanishes at the horizon but its first derivative $a'$ does not, as can be seen from equation (\ref{At}) in Appendix B.

The last important quantity is the entropy density, which is computed by direct use of Hawking formula. This is just
\be s=\frac{4\pi w}{16\pi G_{N}}. \label{eqqents} \ee
Notice that $v$ and $w$ in these expressions come from the metric elements. Using the identification  (\ref{identification}), we see that
\be \frac{e}{v}=a'(r_{H}),  \label{releva}\ee
so, by replacing this in the charge density and using the attractor equation (\ref{QAd}), the charge density is related to the quantity $\tilde{Q}$ in Sen's entropy function as
\be \rho=\tilde{Q}.  \label{relQrho2}\ee
We note that this result coincides with (\ref{relQrho}). This provides a consistency check on the approach presented above.

We see that we can obtain the $AdS_{2}$ factor in the near horizon region only when the linear term in the Taylor expansion (\ref{taylor}) vanishes, i.e. $\nU'(r_{H})=0$. As the temperature of the black hole is directly proportional to this factor, $T\sim \nU'(r_{H})$, this requires that the black hole be extremal in order to obtain the $AdS_{2}\times \R^{2}$ near horizon geometry. For the planar AdS-RN black hole reviewed in section \ref{sec:AdSRN}, the extremality can be achieved even if we set one of the charges to zero, i.e. it is not necessary to have a dyonic black hole to have $T=0$. This can be easily seen by setting the electric or the magnetic charge to zero in the expression for the temperature in (\ref{defgamma}). But this is a feature of the Einstein-Maxwell theory, and it is absent in the EMD theory, in which the coupling to the field strength $Z(\phi)$ is non-constant. For black holes with spherical horizons, as shown in Appendix \ref{sec:appA}, the magnetically charged spherical black hole has a temperature given by (\ref{Htemp}), which is exactly the temperature of the Schwarzschild black hole. The only way to achieve $T=0$ in this case is by having infinite mass $M$, which is physically unacceptable. This means that the non-constant coupling $Z(\phi)$ made the magnetically (or electrically) charged black hole for the EMD theory to have the same structure as the Schwarzschild black hole, with just one horizon and no zero temperature limit, unlike the spherical Reissner-Nordstr\"{o}m black hole, which has two horizons and well defined zero temperature limit. But when the spherical black holes of the EMD theory are dyonic, the black hole has two horizons in the same way as the Reissner-Nordstr\"{o}m black hole. Then,  the temperature is given by (\ref{Temp}), and it is exactly zero if and only if the horizons coincide. In other words, the black holes with spherical horizons for the EMD theories can only be extremal if it contains electric and magnetic charge at the same time, i.e. it is dyonic. We would expect the same behavior for planar black holes, but it turns out that the potential must be non-zero in order to obtain regular planar black holes\footnote{Regularity of the solutions in the extremal limit is guaranteed if the attractor equations admit solutions, since they are also solutions to the equations of motion with $AdS_{2}\times \R^{2}$ near-horizon geometry}. If we turn on the potential, as is the case in this paper, we may have a well defined extremal limit with $AdS_{2}\times \R^{2}$ near-horizon geometry, even for cases in which the black hole contains only electric charge for instance. This is due to the fact that $\nU'(r_{H})$ may be put to zero when the potential is non-trivial. But having $AdS_{2}\times \R^{2}$ near-horizon geometry in the extremal limit does not guarantee that the electric field or dilaton field are non-zero or finite in the extremal limit when evaluated on the horizon. We must obtain the solutions to the attractors to show this explicitly. Since DC conductivities are written in terms of fields evaluated on the horizon of the black hole, its finiteness is directly related to the finiteness of these fields on the horizon. Here, in this paper, we obtain non-trivial and finite fields on the horizon only when the black hole is dyonic, so, the DC conductivities are also non-trivial and finite. This analysis also tells us that the entropy function formalism only fails when the black hole does not contain $AdS_{2}$ factor in the near-horizon limit, or when one of the fields on the horizon is infinite, but both limitations are absent for the dyonic black holes considered in the paper.

With these results, we are now able to compute conductivities explicitly for different theories. We will analyze three models separately. First we will write the solution to the attractor equations and then we combine with the expressions for the conductivities (\ref{elconddxx}), (\ref{elconddxy}), (\ref{etconddxx}) and (\ref{etconddxy}). 

\subsection{Massless scalar\label{sec:MS}}
In this case the potential assumes the form
\be V(\phi)=-\frac{6}{L^2}, \ee
where we wrote $V(\phi)$ only to keep notation, since the potential is a constant and does not really depend on the dilaton field $\phi$. Equation (\ref{rank}) gives a map between the number of colors in the gauge theory $N$ and the constants in the gravity theory, i.e. Newton's constant $G_{N}$ and the $AdS_{4}$ scale $L$. As the constant potential satisfies the requirement given by equations (\ref{PotentialCond}),  the EMD theory with such potential is  dual to a three-dimensional conformal field theory. Although this is the simplest potential one can consider, to the best of our knowledge, a full dyonic black hole solution for this theory is not known, so writing explicitly the conductivities of the previous section in terms of the black hole parameters at finite temperature is not possible. In this subsection, we show that it is possible to solve the attractor equations for this theory and consequently for the extremal case, we can write the conductivities explicitly in terms of the black hole parameters for the extremal case. Moreover, we also use equation (\ref{rank}) and write the thermoelectric conductivity $\alpha_{xy}$ in terms of the rank of the gauge group $N$. 

The attractor equations for this theory admit the solution
\be Z(u_{D})=g_{4}^{2}\frac{\tilde{Q}}{B}, \,\,\, e=\sqrt{\frac{L^{2}B}{6(16 \pi G_{N})\tilde{Q}}}, \,\,\, v=\frac{L^{2}}{6}, \,\,\, 
w=\sqrt{\frac{L^{2}(16 \pi G_{N})\tilde{Q}B}{6}} . \label{solmassless}\ee
The solution is independent of the functional form of the coupling $Z(u_{D})$. The entropy density follows from the solution for $w$,
\be s=4\pi\sqrt{\frac{L^{2}}{6}\frac{\tilde{Q} B}{(16\pi G_{N})}}=\frac{1}{3\gamma}\sqrt{\frac{\mathcal{Q}B}{3}}\,N^{3/2}\,, \label{entropyMS}\ee
where we defined $\mathcal{Q}\equiv g_{4}^{2}\rho$ as the normalized charge density of the black hole and we used (\ref{rank}) to rewrite $s$ in order to show explicitly the scaling of this result with the rank $N$ of the gauge group. Note that $\mathcal{Q}$ is independent of $N$, indeed we have
\be \rho=\tilde{Q}=\frac{\mathcal{Q}}{g_{4}^{2}}=\mathcal{Q}\frac{2L^2}{\gamma^2\kappa_{4}^{2}}=\frac{\sqrt{2}\mathcal{Q}}{6\pi\gamma^2}\,N^{3/2}\,. \ee
Replacing these results in (\ref{elconddxx}), (\ref{elconddxy}), (\ref{etconddxx}) and (\ref{etconddxy}) we arrive at our first results,
\be \sigma_{xx}=0, \,\,\, \sigma_{xy}=\frac{\tilde{Q}}{B}=\frac{\sqrt{2}\mathcal{Q}}{6\pi\gamma^2 B}\,N^{3/2}, \label{CosmConstCond}\ee
\be \alpha_{xx}=0, \,\,\,  \alpha_{xy}=4\pi\sqrt{\frac{\tilde{Q}}{6B}\frac{L^{2}}{16 \pi G_{N}}}=\frac{1}{3\gamma}\sqrt{\frac{\mathcal{Q}}{3B}}\,N^{3/2} . \label{thermalN}\ee
Notice that we just rewrote (\ref{CosmConstCond}) for completeness, since this is a general result for EMD theories. The thermoelectric conductivity at zero temperature (\ref{thermalN}) for constant potential is obtained for the first time here. We note that both thermoelectric and Hall conductivities scale as $N^{3/2}$, as is generically expected for theories in 2+1 dimensions  \cite{Hartnoll:2007ai}.

If the dilaton is set to zero\footnote{The same argument applies if we set the dilaton to a constant, instead of zero. We consider only the vanishing dilaton case here for simplicity.} in the EMD theory that is being considered in this section, then these conductivities should match the results computed for the AdS-RN black hole (\ref{condAdSRN}), since the potential used here matches the cosmological constant of $AdS$ (it is the same potential as in section \ref{sec:AdSRN}).  In supergravity theories, the only other non-trivial coupling, $Z(\phi)$, is generically an exponential of the type $Z(\phi)=e^{{\tilde{\gamma}} \phi}$ for an arbitrary constant ${\tilde{\gamma}}$, so it reduces to $Z(0)=1$ if the dilaton is set to zero. This comparison with section \ref{sec:AdSRN} can be used as a consistency check on our new results.

However, notice that in (\ref{condAdSRN}) the entropy density is written in terms of the charges, so in order to make a comparison, we first need to express explicitly the conductivities of the extremal AdS-RN case in terms of the charges too. Precisely, \cref{eqqrho,eqqents} give the corresponding expressions for $s$ and $\rho$, which can be evaluated for the horizon data of (\ref{solutionattr3}), leading to the non-trivial result
\be \sigma_{xy}=\frac{\tilde{Q}}{B}=\frac{\sqrt{2}\mathcal{Q}}{6\pi\gamma^2 B}\,N^{3/2}, \label{AdS-RNCond}\ee
\be \alpha_{xy}=\frac{s}{B}=\frac{4\pi}{B}\frac{L^{2}}{2\kappa^{2}}\sqrt{\frac{1}{3\gamma^{2}}\left(B^{2}+g_{4}^{4}\rho^{2}\right)}=\frac{N^{3/2}}{3B\sqrt{2}\gamma}\sqrt{\frac{1}{3}\left(B^{2}+\mathcal{Q}^{2}\right)}. \label{AdS-RNTCond}\ee
On the other hand, if we set $\phi=0$, then the first equation in (\ref{solmassless}) implies the constraint
\be \mathcal{Q}=B, \label{QB}\ee
which can be inserted into \cref{AdS-RNCond,AdS-RNTCond} to give
\be \sigma_{xy}=\frac{\sqrt{2}}{6\pi\gamma^2}\,N^{3/2}, \,\,\, \alpha_{xy}= \frac{N^{3/2}}{3\sqrt{3}\gamma}. \ee
Indeed, the same results are obtained by inserting (\ref{QB}) into the results of this section, \cref{CosmConstCond,thermalN}. In general terms, the EMD theory reduces to Einstein-Maxwell theory (whose black hole solution is the AdS-RN solution) with an extra constraint on the charges, eq. (\ref{QB}), which results from the dilaton equation of motion in the $\phi=0$ case. From this analysis, it is natural to assume that the conductivities of the EMD theory scales with the same powers of $N$ as in the case of the conductivities of Einstein-Maxwell theory.

\subsection{Exponential couplings\label{subsec:exp}}
This model is defined by\footnote{The parameter $\gamma$ here is an arbitrary constant. This must not be confused with the parameter $\gamma$ of equation (\ref{defgamma}), which will not appear in the remaining part of this paper.}
\be Z(\phi)=e^{\gamma \phi}, \,\,\, V(\phi)=2\beta e^{-\delta \phi} \, ,
\label{exponentials}
\ee
which corresponds to a Liouville potential. Black holes with cylindrical symmetry for this theory were considered in \cite{Charmousis:2009xr}.  This theory was also considered  on page 23 in \cite{Charmousis:2010zz} for electrically charged black holes. For black holes with spherical horizons, the analysis done so far were all numerical: there are no analytical result for conductivities of these theories. As stated before, we will express the conductivities here analytically for the zero temperature case. 

As $V'(0)\neq 0$,  equations (\ref{PotentialCond}) are not fully satisfied, so this system does not admit AdS$_4$ vacuum solution. However, there are some interesting examples of top-down theories that are very difficult to treat analytically, even with the present formalism. For instance, if $V(\phi) = - \frac{6}{L^2} \cosh{(\phi/\sqrt{3})},~ Z(\phi) = {1}/{\cosh{(\phi\sqrt{3}})}$, which is a string-theory inspired model, the attractor equations cannot be solved analytically due to the high powers of the variables in the algebraic equations. The example of this subsection is used as an approximation of these theories, provided that the value of scalar field on the horizon is large, so that the hyperbolic function can be approximated by an exponential. We shall obtain results for exponential potential and coupling and use them for cases in which the condition of large values for the scalar is satisfied. 

In this model, $\gamma$ and $\delta$ are parameters defining the theory, and the constant $\beta$ must be related to the AdS$_4$ radius, as will be done in the end of this subsection. As discussed
in \cite{Charmousis:2009xr,Charmousis:2010zz},  the case
 $\gamma \delta =1$ is of special interest since the associated models arise within string theory.
  The formulae we write are valid for general values of $\gamma$ and $\delta$. Using (\ref{QAd}) and (\ref{derphid}) we have
\be \frac{e}{v}=\frac{g_{4}^{2}\tilde{Q}}{we^{\gamma u_{D}}}, \ee
\be \frac{e^{2}}{v^{2}}=\frac{B^{2}}{w^{2}}-\frac{4\beta\delta}{\gamma}\frac{g_{4}^{2}}{(16 \pi G_{N})}e^{-(\delta+\gamma)u_{D}}.  \ee
Replacing the last equation in (\ref{dervd}) we have
\be \frac{B^{2}}{w^{2}}=-2\beta\frac{g_{4}^{2}}{(16 \pi G_{N})}\left(1-\frac{\delta}{\gamma}\right)e^{-(\delta+\gamma)u_{D}}. \ee
Combining these three equations, we obtain as solution to the attractor equations
\be e^{u_{D}}=\left[\frac{g_{4}^{4}\tilde{Q}^{2}}{B^{2}}\frac{(\gamma-\delta)}{(\gamma+\delta)}\right]^{\frac{1}{2\gamma}}, \label{scalaremdh}\ee
\be e=\frac{g_{4}^{3}\tilde{Q}}{B}\sqrt{\frac{(\gamma-\delta)}{-2\beta(16\pi G)\gamma}}\left[\frac{g_{4}^{4}\tilde{Q}^{2}}{B^{2}}\frac{(\gamma-\delta)}{(\gamma+\delta)}\right]^{\frac{\delta-3\gamma}{4\gamma}},\label{elfieldemdh} \ee
\be v=-\frac{1}{2\beta}\left[\frac{g_{4}^{4}\tilde{Q}^{2}}{B^{2}}\frac{(\gamma-\delta)}{(\gamma+\delta)}\right]^{\frac{\delta}{2\gamma}}, \label{vemdh}\ee
\be w=B\sqrt{\frac{(16\pi G)\gamma}{-2\beta g_{4}^{2}\left( \gamma-\delta\right)}}\left[\frac{g_{4}^{4}\tilde{Q}^{2}}{B^{2}}\frac{(\gamma-\delta)}{(\gamma+\delta)}\right]^{\frac{\delta+\gamma}{4\gamma}}. \label{wemdh}\ee
The entropy density is given by
\be s=4\pi B\sqrt{\frac{\gamma}{-2\beta g_{4}^{2}(16\pi G)\left( \gamma-\delta\right)}}\left[\frac{g_{4}^{4}\tilde{Q}^{2}}{B^{2}}\frac{(\gamma-\delta)}{(\gamma+\delta)}\right]^{\frac{\delta+\gamma}{4\gamma}}.  \ee
{ Although the simple relation (\ref{rank}) does not readily work in this set-up we can provide an argument that a similar dependence on the rank of the gauge group holds for that case. First, let us observe that for dimensional reasons the full potential should  have form 
\begin{equation}
V_{f}(\phi) = - \frac{6}{L^2_{\text{AdS}_4}}f(\phi)
\end{equation}
where $f$ is a dimensionless function. Our approximate potential (\ref{exponentials}) has a similar form with dimensionful constant $\beta$ and dimensionless function $\exp(-\delta\phi)$. Then the approximation mentioned  when introducing this model is in fact $f(\phi\vert_{\text{hor}})\approx\exp(-\delta\phi\vert_{\text{hor}})$ for some values of $\phi\vert_{\text{hor}}$ the dilaton at the horizon, which leaves the dimensionful constant unchanged. That leads to the conclusion that $\beta\sim - {1}/{L^2}$, 
}
and the entropy density will have the same dependence in $N$ as in the previous model, i.e.
$ s\propto N^{3/2}$. 
{ It is worth mentioning that in the argument above we only identify the \emph{AdS$_4$ scale} with the parameter of our model, while the AdS$_2$ scale remains dynamically determined. Purely from dimensional reasons, one would expect relation such as
\be L_{\text{AdS}_2} = F\left(\tilde{Q}/B, \ldots\right)L_{\text{AdS}_4} \ee
to hold, with dots standing for the parameters of the model. This is indeed the case, if one compares the above with (\ref{vemdh}) it matches the expectations based on dimensional analysis, and provides further evidence for the argument given above. 
}

For the model of this subsection, we study the cases in which $\delta=1/\gamma$, and the nonzero conductivities of \eqref{eq:rhos} are given by
\be \sigma_{xy}=\frac{\tilde{Q}}{B},\,\,\,\alpha_{xy}=\frac{4\pi \gamma}{\sqrt{-2\beta g_{4}^{2}(16\pi G)\left( \gamma^{2}-1\right)}}\left[\frac{g_{4}^{4}\tilde{Q}^{2}}{B^{2}}\frac{(\gamma^{2}-1)}{(\gamma^{2}+1)}\right]^{\frac{\gamma^{2}+1}{4\gamma^{2}}}.\ee
As discussed in \cite{Charmousis:2010zz}, the cases $\gamma=\sqrt{3}$ and $\gamma=1$ are of special interest: $\gamma=1$ arises from string theory in four dimensions with a vector arising from the NS sector, while $\gamma=\sqrt{3}$ arises from a Kaluza-Klein reduction. 
In our case, for $\gamma=\sqrt{3}$ our conductivity results reduce to
\be \sigma_{xy}=\frac{\tilde{Q}}{B},\,\,\, \alpha_{xy}=\frac{2\pi}{\sqrt{(16\pi G)}}  \sqrt{-\frac{3}{\beta }}\frac{1}{g_{4}^{2/3}2^{1/3}}\left(\frac{\tilde{Q}}{B}\right)^{\frac{2}{3}}.\ee
On the other hand, for $\gamma=1$ we have
\be \sigma_{xy}=\frac{\tilde{Q}}{B},\,\,\, \alpha_{xy}=2\pi\frac{g_{4}}{\sqrt{(16\pi G_{N})}}\frac{\tilde{Q}}{B}\sqrt{-\frac{1}{\beta} }.\ee

\begin{figure}
\begin{subfigure}{0.48\textwidth}
\includegraphics[width=\linewidth]{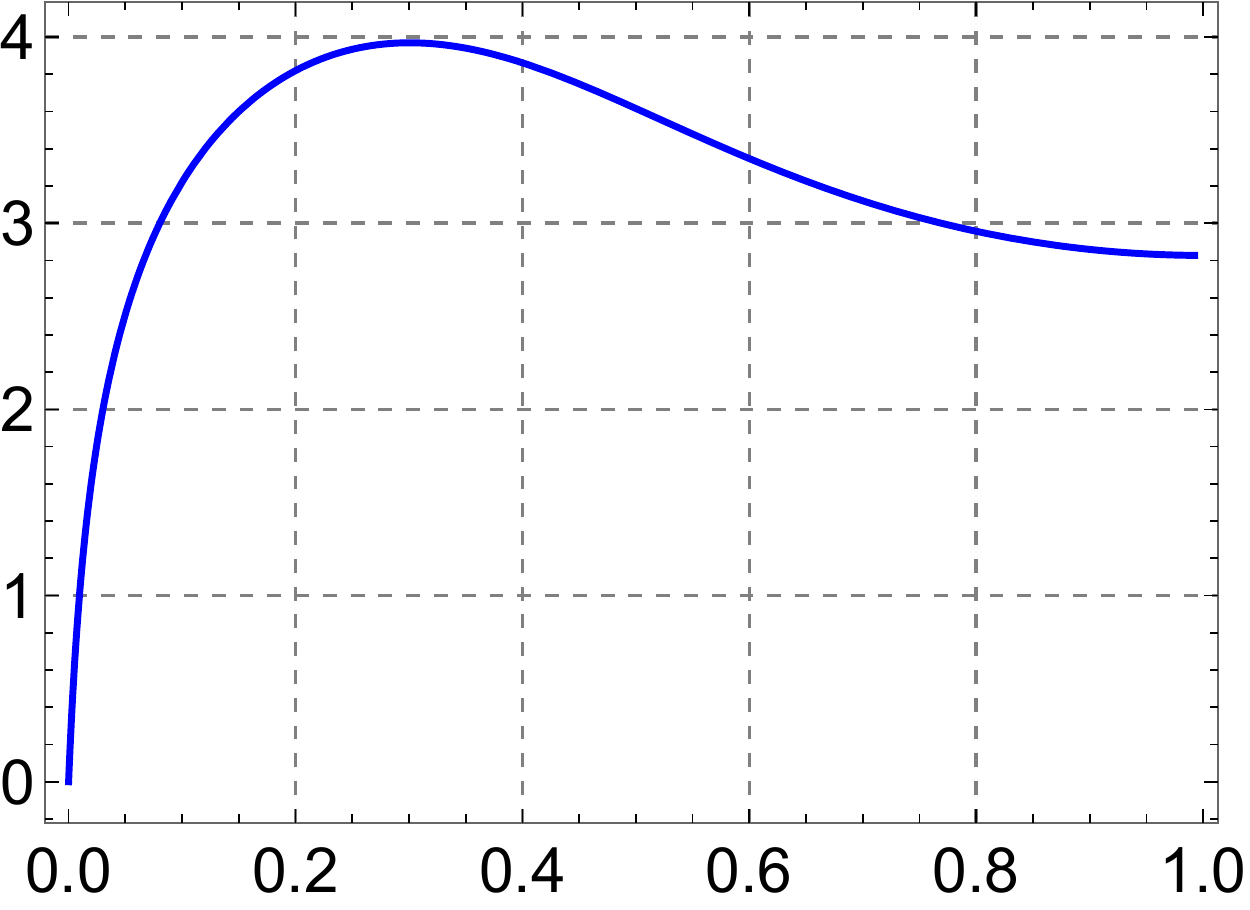}
\caption{Dilaton $\phi$ profile along the bulk.} \label{figphi}
\end{subfigure}
\put(-15,-45){$\tr$}
\put(-218,70){$\phi$}
\hspace*{\fill} 
\begin{subfigure}{0.48\textwidth}
\includegraphics[width=\linewidth]{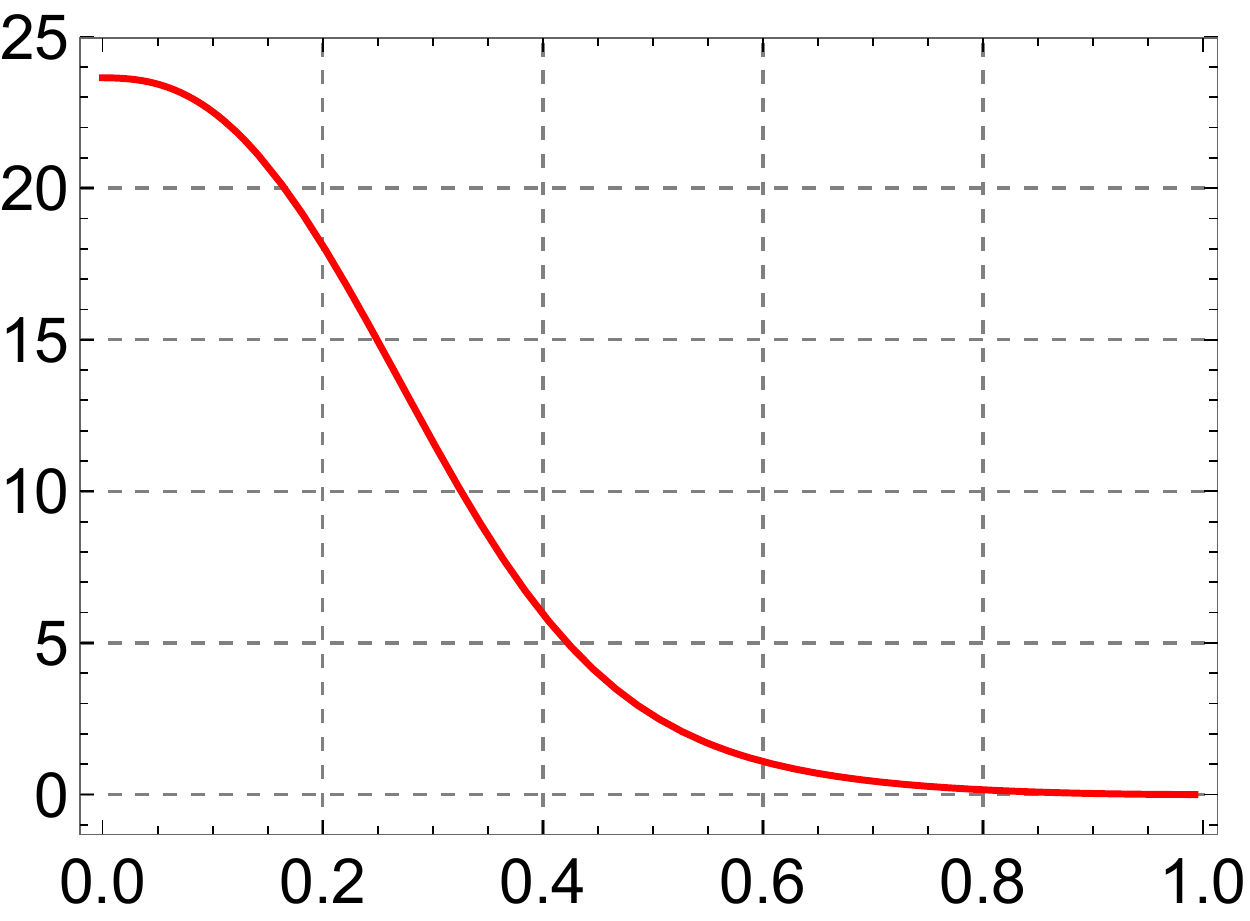}
\caption{Gauge field $A_t$ profile along the bulk.} \label{figat}
\end{subfigure}
\put(-15,-40){$\tr$}
\put(-208,70){$A_t$}
\caption{Result of RG flow for the case $\gamma=-\sqrt{3}$, $\delta=-1/\sqrt{3}$.
In this case, the magnetic field needs to be $B \gg Q$ in order for the
coupling $Z(\phi)$ to be well-approximated by an exponential near the
horizon. We chose $B/Q=100$ for this plot. The horizon is at $\tr=1$, the
boundary at $\tr=0$.} \label{fig:1}
\end{figure}

\begin{figure}
\begin{subfigure}{0.48\textwidth}
\includegraphics[width=\linewidth]{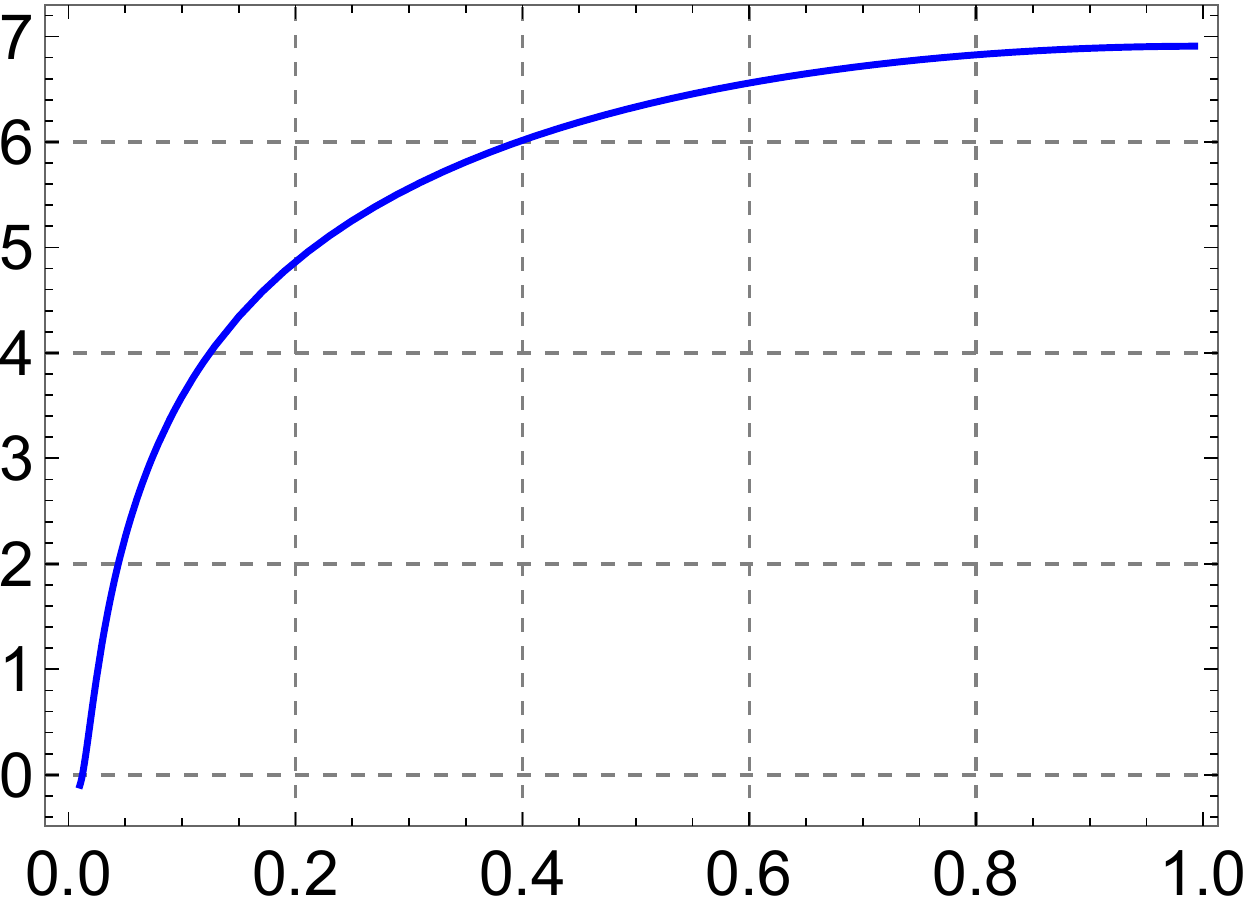}
\caption{Dilaton $\phi$ profile along the bulk.} \label{figphi2}
\end{subfigure}
\put(-2,-35){$\tr$}
\put(-218,70){$\phi$}
\hspace*{\fill} 
\begin{subfigure}{0.48\textwidth}
\includegraphics[width=\linewidth]{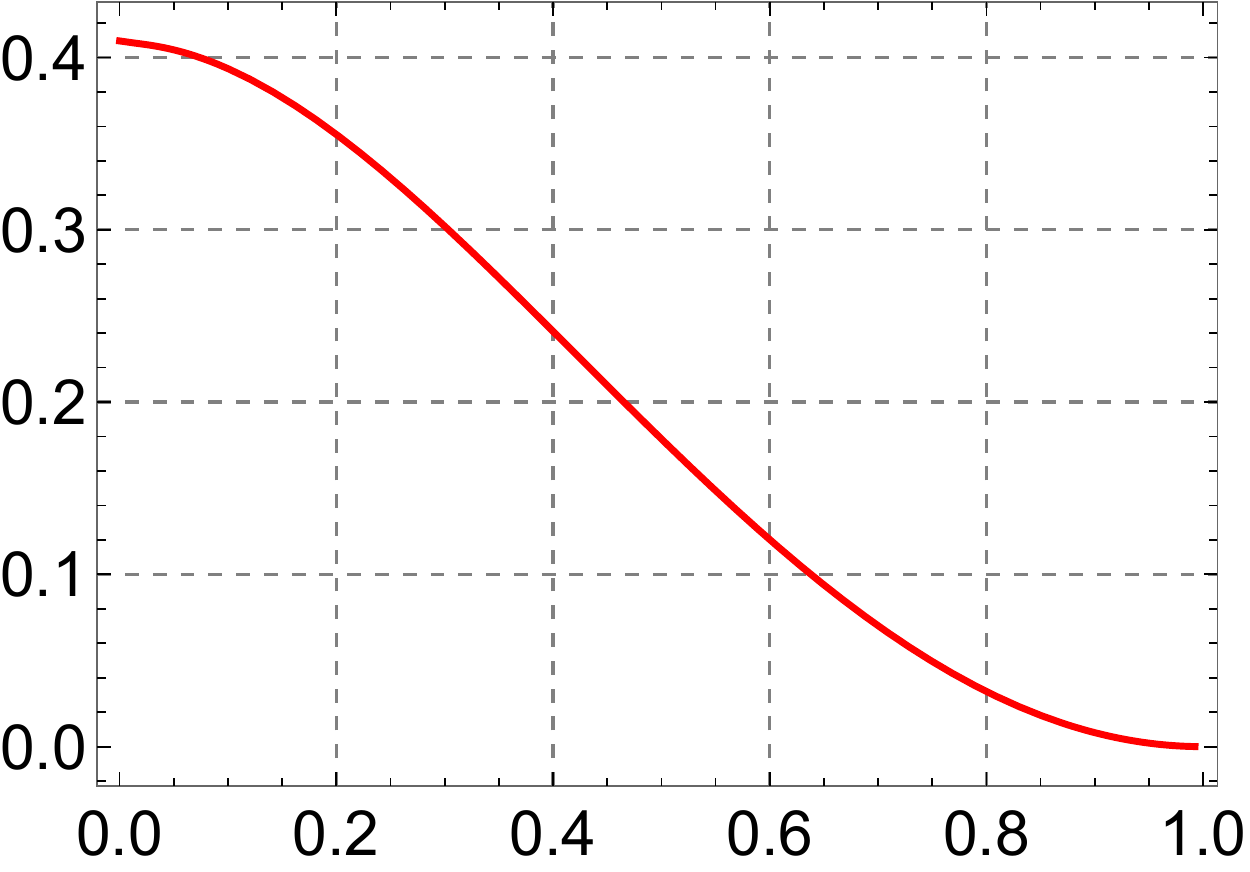}
\caption{Gauge field $A_t$ profile along the bulk.}  \label{figat2}
\end{subfigure}
\put(-2,-33){$\tr$}
\put(-208,85){$A_t$}
\caption{Result of RG flow for the case $\gamma\simeq-1$, $\delta\simeq-1$ and $B/Q=1$. In this case, the magnetic field is not required to be large. The horizon is at $\tr=1$, the boundary at $\tr=0$.} \label{fig:2}
\end{figure}

\noindent For these two special values of $\gamma$, which correspond to top-town models, we numerically computed the bulk solution that connects this solution in the IR to $AdS_4$ in the UV, in order to check that the attractor mechanism provides the appropriate solution. For more details on the numerical UV completion of the solutions, see Appendix \ref{appuv}. Note that the analytical continuation of the coupling and potential (\ref{coshpot}) usually requires us to select appropriate values of $B$ and $Q$ such that $B\gg Q$, in order to produce a large dilaton at the horizon. However, when both $\gamma$ and $\delta$ are close to -1, the dilaton becomes large at the horizon for regular values of the ratio $B/Q$. The plots are presented in Figs. \ref{fig:1} and \ref{fig:2}.

\subsection{Quadratic couplings\label{subsec:quad}}

This is a simple bottom-up model where both potential and gauge coupling are second order polynomials, namely
\be Z(\phi)=1+\frac{\alpha}{2}\phi^{2}, \,\,\, V(\phi)=-\frac{6}{L^{2}}+\frac{\beta}{2L^2}\phi^{2}. \ee
This case was investigated in \cite{Cadoni:2011kv} and may be viewed as an approximation of more complicated top-down models in case where the scalar is small near the horizon. Since the constants are arbitrary, we expect this model to capture universal features. First we note that (\ref{derphid}) reduces to a linear equation in $u_D $, which is the value of the dilaton on the horizon,
\begin{equation}
-\frac{\alpha  u_D  \left(\frac{2 B^2}{w^2}-\frac{2 e^2}{v^2}\right)}{g_4^2}-\frac{8 \beta  u_D}{16\pi G L^2} = 0 \, .
\end{equation}

\noindent Therefore there are two possible solutions. One of them is $u_D = 0$, in which case the other three attractor equations are solved by the dyonic AdS-Reissner-Nordstr\"om geometry without scalar, as in the example of section \ref{sec:AdSRN}. A further solution is obtained by assuming $u_D\neq0$. Equation (\ref{QAd}) and (\ref{derphid}) give
\be \frac{e}{v}=\frac{g_{4}^{2}\tilde{Q}}{wZ(u_{D})}, \label{eqc}\ee
\be \frac{1}{2g_{4}^{2}}\frac{e^{2}}{v^{2}}=\frac{1}{2g_{4}^{2}}\frac{B^{2}}{w^{2}}+\frac{\beta}{\alpha}\frac{1}{(16\pi G)L^{2}}. \label{e2qc}\ee
We use (\ref{e2qc}) to eliminate the term containing $e$ from (\ref{sub}) and write
\be \frac{B^{2}}{g_{4}^{2}w^{2}}Z(u_{D})=\frac{1}{L^{2}(16\pi G)}\left(6-\frac{\beta}{\alpha}\right)-\frac{\beta}{L^{2}(16\pi G)}u_{D}^{2}.  \label{E1qc}\ee
We also use (\ref{eqc}) to eliminate $e$ in (\ref{e2qc}) and write
\be \frac{1}{g_{4}^{2}w^{2}}\left(\frac{g_{4}^{4}\tilde{Q}^{2}}{Z(u_{D})^{2}}-B^{2}\right)=\frac{2\beta}{\alpha}\frac{1}{L^{2}(16\pi G)}. \label{E2qc}\ee
Combining equations (\ref{E1qc}) and (\ref{E2qc}) we can write the quadratic equation for $u_{D}^{2}$,
\be u_{D}^{4}+\frac{u_{D}^{2}}{\alpha}\left(1+\frac{\beta}{(\beta-6\alpha)}\frac{g_{4}^{4}\tilde{Q}^{2}}{B^{2}}\right)+\frac{1}{\alpha^{2}}\left(1+\frac{g_{4}^{4}\tilde{Q}^{2}}{B^{2}}\frac{(\beta-6\alpha)}{(\beta+6\alpha)}\right)=0.  \ee
The solution for this equation is
\be 1+\frac{\alpha}{2}u_{D}^{2}=-\frac{g_{4}^{4}\tilde{Q}^{2}}{B^{2}}\frac{\beta}{(3\alpha-2\beta)}\pm \sqrt{-3+\frac{g_{4}^{2}\tilde{Q}^{2}}{B^{2}(\beta+6\alpha)}\left((24\alpha-6\beta)+\frac{g_{4}^{2}\tilde{Q}^{2}}{B^{2}}\frac{\beta^{2}}{(6\alpha+\beta)}\right)}. \label{solapp}\ee
This is the solution to $u_{D}$ written in terms of the charges, found by solving the attractor equations. The expressions for $w$, $v$ and $e$ written in terms of the charges as provided by the attractor equations are rather lengthy, and we don't write them explicitly here. Instead, we write them in terms of $u_{D}^{2}$, i.e.
\be e=\frac{v}{w}\frac{g_{4}^{2}\tilde{Q}}{(1+\frac{\alpha}{2}u_{D}^{2})}, \ee
\be v=\frac{L^{2}}{(6-\frac{\beta}{2}u_{D}^{2})}, \ee
\be w=\sqrt{\frac{L^{2}(16\pi G)B^{2}\alpha}{g_{4}^{2}} \frac{(1+\frac{\alpha}{2}u_{D}^{2})}{6\alpha-\beta(1 +\alpha u_{D}^{2})}}. \ee
The entropy density is given by
\be s=4\pi B\sqrt{\frac{L^{2}\alpha}{g_{4}^{2}(16\pi G)} \frac{(1+\frac{\alpha}{2}u_{D}^{2})}{6\alpha-\beta(1 +\alpha u_{D}^{2})}}. \ee
From this result we express the conductivities in terms of the black hole parameters as
\be \sigma_{xx}=0, \,\,\, \sigma_{xy}=\frac{\tilde{Q}}{B}, \ee
\be \alpha_{xx}=0, \,\,\,  \alpha_{xy}=4\pi\sqrt{\frac{L^{2}\alpha}{g_{4}^{2}(16\pi G)} \frac{(1+\frac{\alpha}{2}u_{D}^{2})}{6\alpha-\beta(1 +\alpha u_{D}^{2})}}, \ee
where again $u_{D}$ is obtained from equation (\ref{solapp}).

It is natural to ask about the stability of these solutions. In order to tackle this question we perform a simple analysis:  It is likely that possible instabilities are triggered by the scalar sector. We therefore calculate the equation for scalar fluctuations on the dyonic AdS-RN background, and then  compare the effective mass to the Breitenlohner-Freedman (BF) bound. Since the near-horizon geometry is translationally invariant in $t,~x,~y$ directions, we investigate an Ansatz for the scalar of the form 
\begin{equation}
\phi(r, t, x,y) = R(r)e^{- i \omega t + k_1 x + k_2 y}
\end{equation}
subject to 
\begin{equation}
\square_{AdS_2\times\mathrm{R}^2} \phi -\alpha \phi - \frac{2 \kappa^2}{4 g_4^2}\frac{\beta}{L^2} F_{\mu\nu}F^{\mu\nu} \phi= 0.
\end{equation}
From this equation, following the logic of \cite{Moroz:2009kv}, we extract the effective mass generated by the scalar potential and interaction with the background electromagnetic field,
\begin{equation}
m_{\text{eff}}^2 = \frac{B^2 (6 \alpha +\beta )+g_4^4 \tilde{Q}^2 (\beta -6 \alpha )}{6\left(B^2+g_4^4\tilde{Q}^2\right)}.
\end{equation}
The BF bound states that geometry is unstable against scalar perturbations if their mass satisfies:
\begin{equation}
m_{\text{eff}}^2 \leq -\frac{1}{4}.\label{BFAsD2}
\end{equation}
Two consequences follow from this: First, the possible phase transition is driven by the electric field in the bulk, which is dual to the chemical potential of the boundary theory, while the magnetic field tends to stabilize the uncondensed phase by increasing the effective mass. Second, this instability does not occur for all models (i.e not for all values of $\alpha,~\beta$): In the  limit
\begin{equation}
\lim\limits_{\tilde{Q}\rightarrow \infty} m^2_{\text{eff}} = \frac{1}{6} (\beta -6 \alpha ),
\end{equation}
a violation of the BF bound (\ref{BFAsD2}) requires that
\begin{equation}
6 \alpha -\beta > \frac{3}{2}. \label{ViolationCond}
\end{equation}
If this is satisfied, the instability occurs at 
\begin{equation}
\tilde{Q}^2 = \frac{B^2 (6\alpha + \beta+ \frac{3}{2})}{g_4^2(6\alpha - \beta-\frac{3}{2})}.
\end{equation}
This superficial analysis indicates that the model may exhibit a possible quantum critical phase transition (i.e. a phase transition at $T=0$). Instabilities due to fluctuations in the non-scalar sector are also conceivable. It is also not clear what happens if the condition (\ref{ViolationCond}) is not satisfied. This question would require a more sophisticated analysis which we leave for future work. It would also be interesting to investigate the possible connections with the work of \cite{Cadoni:2011kv} and \cite{Cadoni:2009xm},  where the phase structure of similar models is investigated at finite temperature and without magnetic field.

\subsection{Heat conductivity $\kappa/T$ \label{sec:kT}}

The approach presented in this paper may also be applied to the heat conductivity $\kappa$. In particular, we confirm the result of \cite{Grozdanov:2015djs} in which a bound on $\kappa/T$ was derived. In that paper, it was argued that the heat conductivity is always non-zero at finite temperature, so long as the dilaton potential is bounded from below. By applying the method advocated in this paper,  we also  find finite results for $\kappa/T$ even for $T \rightarrow 0$. 

In order to obtain the ratio $\kappa/T$ within the formalism considered here, we take $\Phi=0$ in (\ref{hconddxx}). This gives
\be \frac{\bar{\kappa}_{xx}}{T}=\frac{s^{2}Z}{g_{4}^{2}\left(\rho^{2}+\frac{B^{2}Z^{2}}{g_{4}^{4}}\right)}. \label{kTxx}\ee
In the same way, we  we take $\Phi=0$ in (\ref{hconddxy}), and write the ratio
\be \frac{\bar{\kappa}_{xy}}{T}=\frac{\rho}{B}\frac{s^{2}}{\left(\rho^{2}+\frac{B^{2}Z^{2}}{g_{4}^{4}}\right)}. \label{kTxy}\ee
For general EMD theory as the one given by (\ref{actionEMD}), the dimensionless ratio $\kappa/T$ is strictly finite at finite temperature ($T>0$) if the dilaton potential $V(\phi)$ is bounded from below \cite{Grozdanov:2015djs}. A question that remains to be answered is whether this ratio continues to be finite as $T\rightarrow 0$. Here, we assume that this is the case, and then write $\kappa/T$ for the theories studied in subsections \ref{sec:MS} and \ref{subsec:exp}. We do not write the results for $\kappa/T$  for the model of subsection \ref{subsec:quad} explicitly, since the resulting expressions are lengthy and not so enlightening. However they can be easily found following the same procedure.

\begin{itemize}
\item Massless scalar:
For the massless scalar, the term inside the parentheses in the denominator of (\ref{kTxx}) and (\ref{kTxy}) can be simplified using the solution to the attractor equations (\ref{solmassless}), and the ratio $\kappa/T$ is written as
\be \frac{\bar{\kappa}_{xx}}{T}=\frac{\bar{\kappa}_{xy}}{T}=\frac{s^{2}}{2\rho B}, \ee
where $s$ and $\rho$ are the entropy and charge densities at zero temperature, which are given by (\ref{solmassless}) and (\ref{entropyMS}). Surprisingly,  after using the solution to the attractor equations (\ref{solmassless}),  the values for the ratios $\bar{\kappa}_{xx}/T$ and $\bar{\kappa}_{xy}/T$ coincide at $T=0$,
\be \frac{\bar{\kappa}_{xx}}{T}=\frac{\bar{\kappa}_{xy}}{T}= \frac{(4\pi)^{2}}{12}\frac{L^{2}}{(16\pi G)}=\frac{\pi}{9\sqrt{2}}N^{3/2}. \ee
Unlike the electric and thermoelectric conductivities (\ref{CosmConstCond}) and (\ref{thermalN}), the final result does not depend on the electric and magnetic charges of the black hole.

\item Exponential couplings: For the exponential coupling model, we take $\gamma\delta=1$ and use (\ref{scalaremdh}) to show that
\be \frac{\bar{\kappa}_{xx}}{T}=\frac{s^{2}}{2\rho B}\frac{(\gamma^{2}+1)}{\gamma^{2}}, \,\,\, \frac{\bar{\kappa}_{xy}}{T}=\frac{s^{2}}{2\rho B}\frac{\sqrt{(\gamma^{2}+1)(\gamma^{2}-1)}}{\gamma^{2}}.\ee
Notice that the entropy density squared is given by
\be s^{2}=\frac{(4\pi)^{2} B^{2}\gamma^{2}}{-2\beta g_{4}^{2}(16\pi G)\left( \gamma^{2}-1\right)}\left[\frac{g_{4}^{4}\tilde{Q}^{2}}{B^{2}}\frac{(\gamma^{2}-1)}{(\gamma^{2}+1)}\right]^{\frac{\gamma^{2}+1}{2\gamma^{2}}},  \ee
and it is non-zero for $\gamma=1,\,\sqrt{3}$. Different from the massless scalar case, the ratios $\bar{\kappa}_{xx}/T$ and $\bar{\kappa}_{xy}/T$ for this model are not the same at $T=0$. For $\gamma=\sqrt{3}$ we have
\be \frac{\bar{\kappa}_{xx}}{T}= \frac{(4\pi)^{2}}{(-2\beta)(16\pi G)}\left(\frac{g_{4}^{2}\rho}{4B}\right)^{1/3}, \,\,\,  \frac{\bar{\kappa}_{xy}}{T}=\frac{\sqrt{2}(4\pi)^{2}}{(-2\beta)(16\pi G)}\left(\frac{g_{4}^{2}\rho}{4B}\right)^{1/3}. \ee
For $\gamma=1$, we have 
\be \frac{\bar{\kappa}_{xx}}{T}=\frac{(4\pi)^{2}}{(-2\beta)(16\pi G)}\frac{g_{4}^{2}\rho}{2B}, \,\,\, \frac{\bar{\kappa}_{xy}}{T}=0.\ee
\end{itemize}

We note that the numerical values we obtain for $\kappa/T$ differ from those obtained in \cite{Grozdanov:2015djs}, since in that paper, $L^2/\kappa^2 \propto N^2$ is used instead of \eqref{rank} which is the appropriate expression for a 2+1-dimensional boundary theory.

\subsection{S-duality of the attractor equations and conductivities\label{sec:appC}}

In \cite{Donos:2015bxe}, the authors derive conductivities from a set of generalized Stokes equations at the horizon. For a 3+1-dimensional gravity dual,
 these equations are invariant under S-duality. Here we examine the S-duality transformation properties of the attractor equations considered in the sections above,
restricting our attention to theories whose potential is even and which satisfy
$Z(-\phi) = Z(\phi)^{-1}$. 

The S-duality transformations are given by
\be F^{\mu\nu}\rightarrow Z(\phi)\frac{{\tilde{\epsilon}}^{\mu\nu\rho\sigma}}{2\sqrt{-g}}F_{\rho\sigma}, \,\,\, \phi \rightarrow -\phi, \ee
where ${\tilde{\epsilon}}^{\mu\nu\rho\sigma}$ is the totally antisymmetric Levi-Civita symbol, with ${\tilde{\epsilon}}^{trxy}=1$. If we focus on horizon of the black hole, then the S-duality transformation results in 
\be -\frac{e}{v^{2}}\rightarrow -\frac{1}{vw}Z(u_{D})B, \,\,\, \frac{B}{w^2}\rightarrow -\frac{1}{vw}Z(u_{D})e, \,\,\, Z(u_{D})\rightarrow \frac{1}{Z(u_{D})}. \label{c2}\ee
The S-duality transformation of the charge density $\rho$ is obtained from its definition within EMD theory, from which we obtain
\be   \rho=\sqrt{-g}\frac{Z(\phi)}{g_{4}^{2}}
F^{tr} \rightarrow \sqrt{-g}\frac{1}{Z(\phi)g_{4}^2}Z(\phi)\frac{{\tilde{\epsilon}}^{trxy}}{\sqrt{-g}}F_{xy}=\frac{B}{g_{4}^2}. \label{c3} \ee
Using $\tilde{Q}=\rho$ we see directly that the transformation for the magnetic field in (\ref{c2}) is given by $B\rightarrow - g_{4}^2\rho$.  Applying the transformations (\ref{c2}) and (\ref{c3}) to the attractor equations, we see that (\ref{dervd}), (\ref{derwd}) and (\ref{derphid}) are invariant, and (\ref{QAd}) gives a trivial identity. This shows that the attractor equations are invariant under S-duality transformations.  However note that S-duality is an invariance of the equations of motion, but not of the action, so we may expect that the entropy function \eqref{entrfct} is not invariant under S-duality either. In fact, as was pointed out in reference \cite{Sen:2007qy}, applying an S-duality transformation gives rise to a new entropy function in the attractor formalism, which might generate new attractor equations. Nevertheless, the extremization of both the initial and the S-dual entropy functions yields the same black hole entropy, which shows that the black hole entropy is invariant under S-duality.

The result (\ref{entropyMS})  for the entropy density contains a square root of the product of the charges. This comes from the fact that the attractor equations involve squares. For investigating the S-duality properties of our explicit expressions for the entropy and for the conductivities, we have to take care of the signs carefully when taking square roots of quadratic expressions. As an example, the result for the thermoelectric conductivity of \eqref{thermalN} should be written as
\begin{gather}
\alpha_{xy} = \mathrm{sgn} (B) \, 4 \pi \, \sqrt{\frac{L^2|\tilde{Q}| } {6 |B|} 16 \pi G } \, .
\end{gather}
Then, the transformed conductivities are  given by 
\be \sigma_{xy}\rightarrow -\frac{B}{g_{4}^{4}\tilde{Q}}, \,\,\, \alpha_{xy}\rightarrow -\frac{4\pi }{g_{4}^{2}\tilde{Q}} \sqrt{\frac{\tilde{Q}B}{6}\frac{L^{2}}{(16\pi G)}}, \ee
with $\tilde{Q}\rightarrow \rho$. Performing the transformation once more, we see that the conductivities transform as
\be \sigma_{xy}\rightarrow \sigma_{xy}, \,\,\, \alpha_{xy}\rightarrow -\alpha_{xy} \, ,\ee
which is precisely what authors of \cite{Donos:2015bxe} find.\\
The result for $\kappa/T$ is independent of the absolute values of charges for the massless scalar model, but again its sign depends on signs of charges. This implies that after transforming the charges once we have
\be \bar{\kappa}_{xx}\rightarrow \bar{\kappa}_{xx}, \,\,\, \bar{\kappa}_{xy}\rightarrow -\bar{\kappa}_{xy} \, . \ee
Transforming the charges again results in 
\be \bar{\kappa}_{xx}\rightarrow \bar{\kappa}_{xx}, \,\,\, \bar{\kappa}_{xy}\rightarrow \bar{\kappa}_{xy} \, , \ee
which again is consistent with general transformation laws of \cite{Donos:2015bxe}.

\section{Generalizations and conclusions}

In this paper we computed conductivities at zero temperature for EMD theories. The expressions we obtained are written in terms of the extremal black hole parameters, and we showed that the off-diagonal components of the electric and thermoelectric conductivities scale as
\be \sigma_{xy}\sim N^{3/2},\;\;\alpha_{xy}\sim N^{3/2} \ee
for a constant potential, where $N$ is the rank of the gauge group of the conformal field theory dual to the EMD theory. We argued that this should also be the case for theories with other kinds of potential. We briefly discussed that in the $T=0$ limit the EMD presents different phases, which may be related to quantum phase transitions. All the results were obtained by applying Sen's entropy function method in the AdS/CMT context. We also computed $\kappa_{xx}/T$ and $\kappa_{xy}/T$ for EMD theories assuming that these ratios are finite at $T=0$. For a constant potential, $\kappa_{xx}/T$ is equal to $\kappa_{xy}/T$, and they also scale as
\be \frac{\kappa_{xx}}{T}=\frac{\kappa_{xy}}{T}\sim N^{3/2}. \ee

Explicit analytical zero-temperature expressions are expected to be very useful in particular for universality arguments in AdS/CMT, see for instance \cite{Erdmenger:2015qqa,Kim:2016jjk}
and references therein. We expect that the results of this paper may be generalized to more involved 
geometries relevant in that context.

Although we investigated the simplest EMD theories, generalizations of this approach are indeed possible. The two most immediate ones are the following: One direction is to generalize the equations for conductivities (\ref{elconddxx}, \ref{elconddxy}, \ref{etconddxx}, \ref{etconddxy}, \ref{hconddxx}, \ref{hconddxy}) to take into account multiple scalar fields and gauge fields coupled in the non-minimal way. In the context of AdS/CFT it is useful to consider maximally supersymmetric supergravities, and including more scalars and field strengths will guarantee that we can handle the four-dimensional $\mathcal{N}=8$ gauged supergravity. Solving the attractor equations for the planar case,  the conductivities of the three-dimensional CFT can be expressed explicitly in terms of the black hole parameters and of $N$, just as we did in the paper for simpler cases. Another possibility is to generalize the computation of the conductivities to higher-dimensional supergravity theories. This means that the gravity theory will have a Chern-Simons term in odd dimensions, which may complicate the computation. The entropy function method is valid only for even-dimensional gravity theories, but it is possible to perform dimensional reduction down to even dimensions, treat the even dimensional theory as an effective theory, and then compute the horizon data, in the same spirit as \cite{Astefanesei:2011pz}. That would allow to use the entropy function method to obtain the conductivities at zero temperature also for these cases, and in particular for the case most studied in the literature, which is the supergravity dual of four-dimensional $\mathcal{N}=4$ Super Yang-Mills theory. 

Yet another possibility is to extend the superficial analysis of the phase transition considered in subsection \ref{subsec:quad}. It would be interesting to look not only at the scalar instabilities but also at the full linearized EMD system on both backgrounds. Moreover, as we saw in  \ref{subsec:quad}, the scalar instability occurs only for particular values of parameters $\alpha,~\beta$, but the coexistence of phases with and without scalar fields seems to happen also for other values of those parameters. Exploring this phase diagram would probably require knowledge of the full solution, and therefore one would probably be forced to invoke some numerical methods. Apart from this scalar condensation it seems possible that there may be another phase transition in the regime of vanishing magnetic field, which may lead to a scaling geometry in the far IR.

Our results imply that the attractor mechanism is a very important ingredient in the computation of zero temperature conductivities, as well as in the study of quantum phase transitions.

\vspace{10pt}

{\bf{Acknowledgments}}\\
The authors thank Yago Bea, Ren\'{e} Meyer, Horatiu Nastase and Nick Poovuttikul for useful discussions. We also thank the referee of this paper for very constructive suggestions. DF was
supported by an Alexander von Humboldt Foundation fellowship. PG is grateful for the hospitality of the Max-Planck-Institut f\"{u}r  Physik
(Werner-Heisenberg-Institut), where the largest part of this work was carried out. The work of PG is supported by FAPESP grant 2013/00140-7 and 2015/17441-5.

\appendix

\section{Dilatonic black holes\label{sec:appA}}
In the introduction we claimed that it was necessary to consider a dyonic black hole solution in order to obtain a well-defined zero temperature solution. This analysis is based on spherical black holes, and it must also hold for the planar case. The black hole solutions for EMD theory are listed below.
\begin{itemize}
\item Magnetically charged black holes:
\end{itemize}
This solution was obtained by Gibbons and Maeda \cite{Gibbons:1987ps} and later it was also obtained independently by Garfinkle, Horowitz and Strominger \cite{Garfinkle:1990qj}. It is written as:
\be ds^{2}=-\left(1-\frac{2M}{r}\right)dt^{2}+\frac{dr^{2}}{\left(1-\frac{2M}{r}\right)}+r\left(r-\frac{e^{-2\phi_{0}}P^{2}}{M}\right)d\Omega^{2}_{2},  \ee
\be e^{-2\phi}=e^{-2\phi_{0}}\left(1-\frac{e^{-2\phi_{0}}P^{2}}{Mr}\right),  \ee
\be F_{rt}=0,  \ee
\be F_{\theta\phi}=P\sin\theta,  \ee
\be T=\frac{1}{8\pi M}, \label{Htemp} \ee
where $M$ is the mass of the black hole, $P$ is the magnetic charge and $\phi_{0}$ is the asymptotic value of the dilaton at infinity.  The dilaton charge is given by an integral over a two-sphere at spatial infinity\footnote{The dilaton charge $\Sigma$ should not be confused with the two-sphere surface element $d\Sigma^{\mu}$.}
\be \Sigma=\frac{1}{4\pi}\int d\Sigma^{\mu}\nabla_{\mu}\phi=-\frac{e^{-2\phi_{0}}P^{2}}{M}. \ee
Because this is a negative quantity, the dilaton contributes with a long-range, attractive force between black holes.  The surface $r=2M$ is a regular horizon, and the singularity is at $r=\Sigma$. We can obtain the electrically charged solution by applying electric-magnetic  (S-duality) transformations, i.e.
\be F'^{\mu\nu}=\frac{1}{2\sqrt{-g}}e^{-2\phi}\tilde{\epsilon}^{\mu\nu\rho\sigma}F_{\rho\sigma}, \,\,\, \phi'=-\phi, \label{duality}\ee
where $\tilde{\epsilon}^{\mu\nu\rho\sigma}$ is the totally antisymmetric Levi-Civita symbol. The $g_{tt}$ and $g_{rr}$ part of the metric is exactly the same as in the Schwarzschild solution, and consequently, the temperature is given by (\ref{Htemp}). There is no classical zero temperature limit for this solution. The inclusion of the electric charge via $SL(2,R)$ rotation \cite{Shapere:1991ta} gives the same temperature as the magnetically charged solution, so it does not have a well-defined classical zero temperature limit either.

\begin{itemize}
\item Dyonic black holes:
\end{itemize}
The most general dyonic solution was obtained very recently \cite{Goulart:2016cuv}. This solution contains five independent parameters, which generalizes the one that was given in \cite{Kallosh:1992ii},  which contains four independent parameters. The non-extremal solution is written as
\begin{align}
ds^{2}&=-e^{-\lambda}dt^{2}+e^{\lambda}dr^{2}+C^{2}(r)d\Omega^{2}_{2}, \label{genmets} \\
e^{-\lambda}&= \frac{(r-r_{1})(r-r_{2})}{(r+d_{0})(r+d_{1})}, \,\,\, C^{2}(r)=(r+d_{0})(r+d_{1}),\label{metelements}\\
e^{2\phi}&=e^{2\phi_{0}}\frac{r+d_{1}}{r+d_{0}}, \label{gendil} \\
F_{rt}&=\frac{e^{2\phi_{0}}Q}{(r+d_{0})^{2}}, \,\,\, F_{\theta\phi}=P\sin\theta, \label{gauge}\\
\\
T&=\frac{1}{4\pi}\frac{(r_{2}-r_{1})}{(r_{2}+d_{0})(r_{1}+d_{0})},\label{Temp}
\end{align}
with
\begin{align}
d_{0}&=\frac{-(r_{1}+r_{2})\pm\sqrt{(r_{1}-r_{2})^{2}+8e^{2\phi_{0}}Q^{2}}}{2},  \label{reld0} \\ 
d_{1}&=\frac{-(r_{1}+r_{2})\pm\sqrt{(r_{1}-r_{2})^{2}+8e^{-2\phi_{0}}P^{2}}}{2}. \label{reld1}
\end{align} 
This solution contains five independent parameters: the electric charge $Q$, the magnetic charge $P$, the value of the dilaton at infinity $\phi_{0}$, and two integration constants, $r_{1}$ and $r_{2}$. The boundary conditions must be imposed on $r_{1}$ and $r_{2}$.
The singularity is located at $r_S=-d_{0}$ when $d_{0}>d_{1}$, or at $r_S=-d_{1}$ when $d_{1}>d_{0}$. Just like the Reissner-Nordstr\"{o}m black hole, this solution contains an outer and an inner horizon, written as $r_{2}$ and $r_{1}$ respectively, and the temperature (\ref{Temp}) is zero when $r_{1}=r_{2}\equiv r_{0}$. Changing coordinates to $\rho=r-r_{0}$, the extremal solution is then written as
\begin{align}
ds^{2}&=-e^{-2U}dt^{2}+e^{2U}(d\rho^{2}+\rho^{2}d\Omega^{2}_{2}),  \nonumber \\
e^{2U}&=\left(1\pm\frac{\sqrt{2}e^{\phi_{0}}Q}{\rho}\right)\left(1\pm\frac{\sqrt{2}e^{-\phi_{0}}P}{\rho}\right),  \\
e^{2\phi}&=e^{2\phi_{0}}\frac{(\rho\pm\sqrt{2}e^{-\phi_{0}}P)}{(\rho\pm\sqrt{2}e^{\phi_{0}}Q)}, \label{gendilext} \\
F_{rt}&=\frac{e^{2\phi_{0}}Q}{(\rho\pm\sqrt{2}e^{\phi_{0}}Q)^{2}}, \,\,\, F_{\theta\phi}=P\sin\theta.
\end{align}
Here, we have used isotropic coordinates, i.e. $\rho^{2}=x_{1}^{2}+x_{2}^{2}+x_{3}^{2}$, and consequently $d\rho^{2}+\rho^{2}d\Omega^{2}_{2}=d\vec{x}^{2}$. The horizon is located at $\rho=0$. We see that there exist a very well defined zero temperature limit for the dyonic black holes of the EMD theory, which is not the case for the electrically (or magnetically) charged black hole. There also exists a massless limit for the dyonic solution, which is discussed in \cite{Goulart:2016nkv}.

The entropy function method works for extremal BPS and non-BPS black holes, and that is why it is important to consider dyonic black hole solutions, as we did in the text.

\section{Numerical UV completion \label{appuv}}
In this paper, we work under the assumption that the IR solution provided by the attractor equations can be completed towards the UV, in such a way that a full bulk solution displays $AdS_4$ asymptotics. For the case studied in Sec. \ref{subsec:exp}, we provided an explicit check of this statement by computing the full solution numerically. Here we give some details regarding the computation behind that verification, which constructs an RG flow from the IR solution up to the conformal boundary. For the sake of completeness, the codes used to obtain results of this Appendix are available on-line at \url{https://github.com/picalke/Conductivities2016}. The routines were written and executed in Wolfram Mathematica 11.

It is not possible to directly extrapolate the choice (\ref{exponentials}) for the couplings $V(\phi)$, $Z(\phi)$ at the horizon, because exponential couplings do not behave near the boundary in a way compatible with $AdS$ asymptotics. Therefore, in order to obtain a full solution that connects with these expressions, we need to find a way to analytically continue these expressions.
To this end, we consider the following choice of functions:
\be
V(\phi)=4\beta \cosh(-\delta\phi),\;\;\;\;\;Z(\phi)=\frac{1}{2\cosh(-\gamma\phi)}
\label{coshpot}
\ee
When $\delta=-1/\sqrt{3}$ and $\gamma=-\sqrt{3}$, this is the top-down model presented in \cite{Gauntlett2010}. Also, $\beta=-3/2$ is needed in order for the potential to asymptote to the cosmological constant of AdS when the dilaton vanishes. Assuming a large value of the dilaton at the horizon $u_D\ll 1$, they are well approximated by (\ref{exponentials}), as we will see below.

\begin{figure}[t]
\begin{center}
\includegraphics[width=0.52\textwidth]{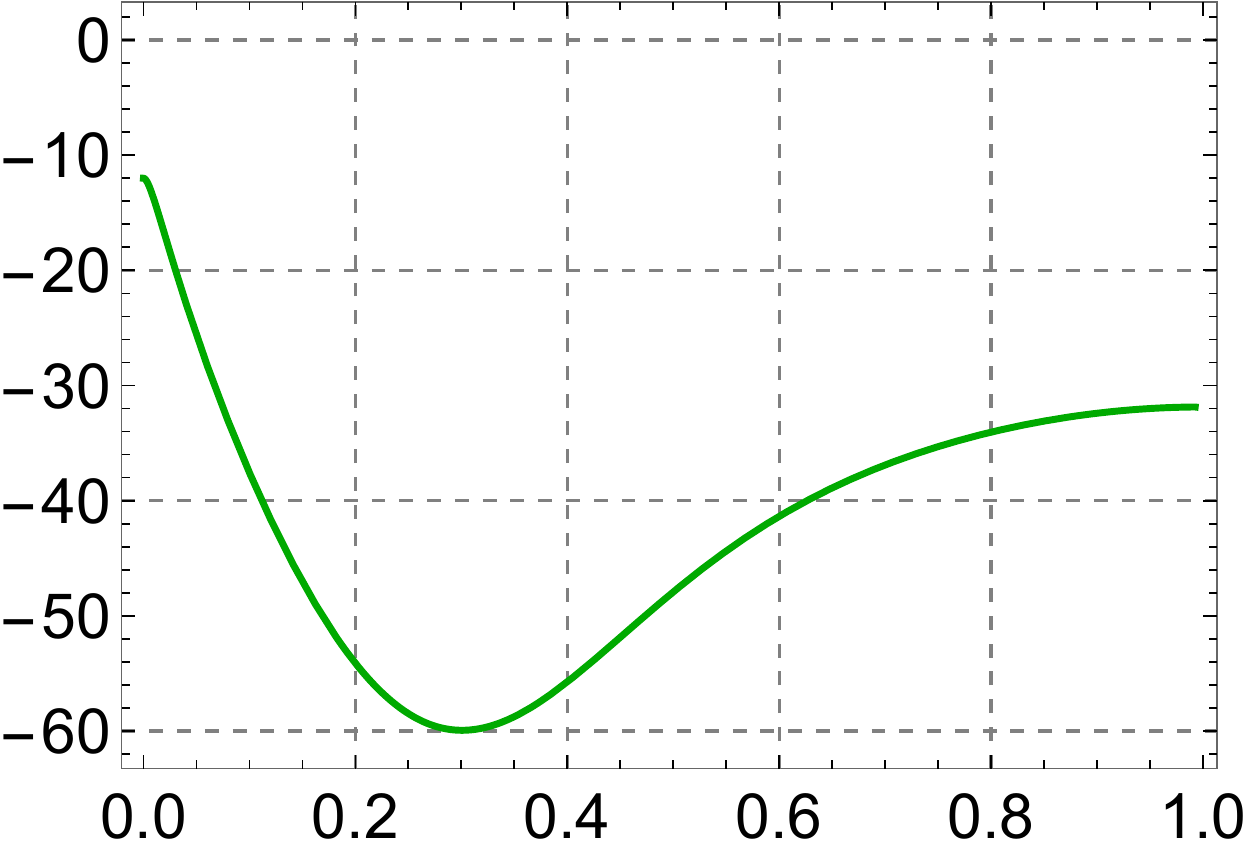}
\put(-2,20){$\tr$}
\put(-228,140){$R$}
\end{center}
\caption{Ricci curvature of the metric \\
for the case $\gamma=-\sqrt{3}$, $\delta=-1/\sqrt{3}$ and $B/Q=100$.} \label{fig:R}
\end{figure}

In order to find the full black hole solution, we adopt the same radial ansatz as in \cite{Donos2016}:
\begin{align}
ds^2&=\frac{1}{g(\tr)^2}\left[-f(\tr)G(\tr)dt^2 + \frac{g'(\tr)^2 G(\tr)}{f(\tr)}d\tr^2 + dx^2+dy^2 \right], \\
A&=A_t(\tr) dt + \frac{B}{2}(x\, dy-y\, dx),\;\;\;\;\phi=\phi(\tr).
\end{align}
Using the gauge freedom of redefining the radial coordinate, $\tr\to R(\tr)$, we can fix the $g_{xx}, g_{yy}$ components by choosing
\be
g(\tr)=\frac{1-(1-\tr)^2}{r_+}\,,
\ee
where $r_+$ is a free constant. The radial coordinate $\tr$ that is fixed by this particular choice relates to the coordinate $\tilde\rho$ in (\ref{adsmetr}) by $\tr=1-\sqrt{\tilde\rho}$ in the near-horizon region. The horizon is located at $\tr=1$ (the conformal boundary lies at $\tr=0$) and in this region, generically $f(\tr)\sim f^{(2)}(\tr-1)^2$. However, the temperature of such a black hole is $T\propto f^{(2)}$. Therefore, in order to find an extremal solution, we impose the expansions
\begin{align}
f(\tr)=f^{(4)}(\tr-1)^4 &+ \mathcal{O}\left((\tr-1)^6\right) \\
G(\tr)=G^{(0)} &+ \mathcal{O}\left((\tr-1)^2\right) \\
A_t(\tr)=A_t^{(2)}(\tr-1)^2 &+ \mathcal{O}\left((\tr-1)^4\right) \label{At}\\
\phi(\tr)=\phi^{(0)} &+ \mathcal{O}\left((\tr-1)^2\right)\,,
\end{align}
where $\phi^{(0)}=u_D$. In order to have $AdS_2\times \mathbb{R}^2$, we need to choose $f^{(4)}=-1/r_+$. The solution is determined by the free parameters $\{\beta,\gamma,\delta,B,r_+\}$. For certain regions of the parameter space $(\gamma,\delta)$, there is a unique regular extremal solution, otherwise no extremal solution can be found. Of course, those are the cases we are interested in\footnote{Exploring the entire parameter regime is beyond the scope of this paper.}. In these cases, all the coefficients of the expansion are fixed by the equations of motion, confirming that the solution is completely determined by the electric charge and magnetic field, as expected from the attractor mechanism.
The other lowest order coefficients are given by
\begin{align}
A_t^{(2)} &= -\sqrt{\frac{\cosh{(\gamma \phi^{(0)})} \sqrt{\gamma+\delta\tanh{(\delta \phi^{(0)})}/\tanh{(\gamma \phi^{(0)})}}}{-2\beta\gamma\cosh{(\delta \phi^{(0)})}}} \\
B^2&=\frac{8r_+^4\beta}{\gamma}\left( \delta\frac{\sinh{(\delta \phi^{(0)})}}{\tanh{(\gamma \phi^{(0)})}}-\gamma\cosh{(\delta \phi^{(0)})} \right) \\
G^{(0)}&=-\frac{1}{4\beta r_+ \cosh{(\delta \phi^{(0)})}}
\end{align}
In order to make a connection to the solution in \cref{scalaremdh,elfieldemdh,vemdh,wemdh}, we need to make the replacement $\cosh{(\delta \phi^{(0)})} \to 1/2 e^{\delta \phi^{(0)}},\; \cosh{(\gamma \phi^{(0)})} \to 1/2 e^{\gamma \phi^{(0)}}$, which is valid as long as $\delta \phi^{(0)}<<0, \gamma \phi^{(0)}<<0$. Also, these expressions are written in terms of the free constant $r_+$, which is related to the charge $Q$ by $r_+^4=\frac{\gamma B^2}{2\beta(\delta-\gamma)}\left(\frac{Q^2}{B^2} \frac{\gamma-\delta}{\gamma+\delta}\right)^{\frac{\delta+\gamma}{2\gamma}}$.

\vspace{10pt}

The equations of motion derived for this ansatz can be integrated numerically using standard shooting techniques, and setting boundary conditions at the horizon. We get four dynamical equations and a constraint. If the former hold, then the constraint equation must be satisfied everywhere provided it is satisfied at the horizon. It serves as a check of the numerics to solve the dynamical equations only and check that the constraint vanishes automatically.

At the boundary, the numerical solution gives a vanishing $\phi$, while the metric asymptotes to
\be
ds^2=\frac{1}{\tr^2}\left[ \frac{\lambda^2 r_+^2}{4}dt^2 + d\tr^2 + \frac{r_+^2}{4}\left(dx^2+dy^2\right) \right],
\ee
where $\lambda$ is a constant determined by the numerical integration which depends non-trivially on the parameters of the solution. Performing the rescalings $\tr\to\frac{\lambda r_+}{2}\hat r$ and $(x,y) \to \lambda (\hat x,\hat y)$, we recognize the familiar form of the $AdS_4$ metric. Such a rescaling corresponds to a dynamically generated $AdS$ scale, a generic feature of holographic RG flows.

\providecommand{\href}[2]{#2}\begingroup\raggedright\endgroup


\begin{thebibliography}{10}

\bibitem{Hartnoll:2009sz}
S.~A. Hartnoll, ``{Lectures on holographic methods for condensed matter
  physics},'' \href{http://dx.doi.org/10.1088/0264-9381/26/22/224002}{{\em
  Class. Quant. Grav.} {\bf 26} (2009)  224002},
\href{http://arxiv.org/abs/0903.3246}{{\tt arXiv:0903.3246 [hep-th]}}.

\bibitem{Hartnoll:2007ai}
S.~A. Hartnoll and P.~Kovtun, ``{Hall conductivity from dyonic black holes},''
  \href{http://dx.doi.org/10.1103/PhysRevD.76.066001}{{\em Phys. Rev.} {\bf
  D76} (2007)  066001},
\href{http://arxiv.org/abs/0704.1160}{{\tt arXiv:0704.1160 [hep-th]}}.

\bibitem{Donos:2015bxe}
A.~Donos, J.~P. Gauntlett, T.~Griffin, and L.~Melgar, ``{DC Conductivity of
  Magnetised Holographic Matter},''
  \href{http://dx.doi.org/10.1007/JHEP01(2016)113}{{\em JHEP} {\bf 01} (2016)
  113},
\href{http://arxiv.org/abs/1511.00713}{{\tt arXiv:1511.00713 [hep-th]}}.

\bibitem{Amoretti:2016cad}
A.~Amoretti, M.~Baggioli, N.~Magnoli, and D.~Musso, ``{Chasing the cuprates
  with dilatonic dyons},''
  \href{http://dx.doi.org/10.1007/JHEP06(2016)113}{{\em JHEP} {\bf 06} (2016)
  113},
\href{http://arxiv.org/abs/1603.03029}{{\tt arXiv:1603.03029 [hep-th]}}.

\bibitem{Erdmenger:2015qqa}
J.~Erdmenger, B.~Herwerth, S.~Klug, R.~Meyer, and K.~Schalm, ``{S-Wave
  Superconductivity in Anisotropic Holographic Insulators},''
  \href{http://dx.doi.org/10.1007/JHEP05(2015)094}{{\em JHEP} {\bf 05} (2015)
  094},
\href{http://arxiv.org/abs/1501.07615}{{\tt arXiv:1501.07615 [hep-th]}}.

\bibitem{Kim:2016jjk}
K.-Y. Kim and C.~Niu, ``{Homes' law in Holographic Superconductor with
  Q-lattices},''
\href{http://arxiv.org/abs/1608.04653}{{\tt arXiv:1608.04653 [hep-th]}}.

\bibitem{Blake:2013bqa}
M.~Blake and D.~Tong, ``{Universal Resistivity from Holographic Massive
  Gravity},'' \href{http://dx.doi.org/10.1103/PhysRevD.88.106004}{{\em Phys.
  Rev.} {\bf D88} (2013) no.~10, 106004},
\href{http://arxiv.org/abs/1308.4970}{{\tt arXiv:1308.4970 [hep-th]}}.

\bibitem{Donos:2014uba}
A.~Donos and J.~P. Gauntlett, ``{Novel metals and insulators from
  holography},'' \href{http://dx.doi.org/10.1007/JHEP06(2014)007}{{\em JHEP}
  {\bf 06} (2014)  007},
\href{http://arxiv.org/abs/1401.5077}{{\tt arXiv:1401.5077 [hep-th]}}.

\bibitem{Donos:2014cya}
A.~Donos and J.~P. Gauntlett, ``{Thermoelectric DC conductivities from black
  hole horizons},'' \href{http://dx.doi.org/10.1007/JHEP11(2014)081}{{\em JHEP}
  {\bf 11} (2014)  081},
\href{http://arxiv.org/abs/1406.4742}{{\tt arXiv:1406.4742 [hep-th]}}.

\bibitem{Blake:2015ina}
M.~Blake, A.~Donos, and N.~Lohitsiri, ``{Magnetothermoelectric Response from
  Holography},'' \href{http://dx.doi.org/10.1007/JHEP08(2015)124}{{\em JHEP}
  {\bf 08} (2015)  124},
\href{http://arxiv.org/abs/1502.03789}{{\tt arXiv:1502.03789 [hep-th]}}.

\bibitem{Lindgren:2015lia}
J.~Lindgren, I.~Papadimitriou, A.~Taliotis, and J.~Vanhoof, ``{Holographic Hall
  conductivities from dyonic backgrounds},''
  \href{http://dx.doi.org/10.1007/JHEP07(2015)094}{{\em JHEP} {\bf 07} (2015)
  094},
\href{http://arxiv.org/abs/1505.04131}{{\tt arXiv:1505.04131 [hep-th]}}.

\bibitem{Sen:2005wa}
A.~Sen, ``{Black hole entropy function and the attractor mechanism in higher
  derivative gravity},''
  \href{http://dx.doi.org/10.1088/1126-6708/2005/09/038}{{\em JHEP} {\bf 0509}
  (2005)  038},
\href{http://arxiv.org/abs/hep-th/0506177}{{\tt arXiv:hep-th/0506177
  [hep-th]}}.

\bibitem{Astefanesei:2011pz}
D.~Astefanesei, N.~Banerjee, and S.~Dutta, ``{Near horizon data and physical
  charges of extremal AdS black holes},''
  \href{http://dx.doi.org/10.1016/j.nuclphysb.2011.07.018}{{\em Nucl. Phys.}
  {\bf B853} (2011)  63--79},
\href{http://arxiv.org/abs/1104.4121}{{\tt arXiv:1104.4121 [hep-th]}}.

\bibitem{Kundu:2012jn}
N.~Kundu, P.~Narayan, N.~Sircar, and S.~P. Trivedi, ``{Entangled Dilaton
  Dyons},'' \href{http://dx.doi.org/10.1007/JHEP03(2013)155}{{\em JHEP} {\bf
  03} (2013)  155},
\href{http://arxiv.org/abs/1208.2008}{{\tt arXiv:1208.2008 [hep-th]}}.

\bibitem{Goldstein:2009cv}
K.~Goldstein, S.~Kachru, S.~Prakash, and S.~P. Trivedi, ``{Holography of
  Charged Dilaton Black Holes},''
  \href{http://dx.doi.org/10.1007/JHEP08(2010)078}{{\em JHEP} {\bf 08} (2010)
  078},
\href{http://arxiv.org/abs/0911.3586}{{\tt arXiv:0911.3586 [hep-th]}}.

\bibitem{Goldstein:2010aw}
K.~Goldstein, N.~Iizuka, S.~Kachru, S.~Prakash, S.~P. Trivedi, and A.~Westphal,
  ``{Holography of Dyonic Dilaton Black Branes},''
  \href{http://dx.doi.org/10.1007/JHEP10(2010)027}{{\em JHEP} {\bf 10} (2010)
  027},
\href{http://arxiv.org/abs/1007.2490}{{\tt arXiv:1007.2490 [hep-th]}}.

\bibitem{Shapere:1991ta}
A.~D. Shapere, S.~Trivedi, and F.~Wilczek, ``{Dual dilaton dyons},''
\href{http://dx.doi.org/10.1142/S0217732391003122}{{\em Mod. Phys. Lett.} {\bf
  A6} (1991)  2677--2686}.

\bibitem{Kallosh:1992ii}
R.~Kallosh, A.~D. Linde, T.~Ortin, A.~W. Peet, and A.~Van~Proeyen,
  ``{Supersymmetry as a cosmic censor},''
  \href{http://dx.doi.org/10.1103/PhysRevD.46.5278}{{\em Phys. Rev.} {\bf D46}
  (1992)  5278--5302},
\href{http://arxiv.org/abs/hep-th/9205027}{{\tt arXiv:hep-th/9205027
  [hep-th]}}.

\bibitem{Cadoni:2011kv}
M.~Cadoni and P.~Pani, ``{Holography of charged dilatonic black branes at
  finite temperature},'' \href{http://dx.doi.org/10.1007/JHEP04(2011)049}{{\em
  JHEP} {\bf 04} (2011)  049},
\href{http://arxiv.org/abs/1102.3820}{{\tt arXiv:1102.3820 [hep-th]}}.

\bibitem{Sen:2007qy}
A.~Sen, ``{Black Hole Entropy Function, Attractors and Precision Counting of
  Microstates},'' \href{http://dx.doi.org/10.1007/s10714-008-0626-4}{{\em
  Gen.Rel.Grav.} {\bf 40} (2008)  2249--2431},
\href{http://arxiv.org/abs/0708.1270}{{\tt arXiv:0708.1270 [hep-th]}}.

\bibitem{Ferrara:1995ih}
S.~Ferrara, R.~Kallosh, and A.~Strominger, ``{N=2 extremal black holes},''
  \href{http://dx.doi.org/10.1103/PhysRevD.52.R5412}{{\em Phys. Rev.} {\bf D52}
  (1995)  R5412--R5416},
\href{http://arxiv.org/abs/hep-th/9508072}{{\tt arXiv:hep-th/9508072
  [hep-th]}}.

\bibitem{Ferrara:1996dd}
S.~Ferrara and R.~Kallosh, ``{Supersymmetry and attractors},''
  \href{http://dx.doi.org/10.1103/PhysRevD.54.1514}{{\em Phys. Rev.} {\bf D54}
  (1996)  1514--1524},
\href{http://arxiv.org/abs/hep-th/9602136}{{\tt arXiv:hep-th/9602136
  [hep-th]}}.

\bibitem{Kunduri:2007vf}
H.~K. Kunduri, J.~Lucietti, and H.~S. Reall, ``{Near-horizon symmetries of
  extremal black holes},''
  \href{http://dx.doi.org/10.1088/0264-9381/24/16/012}{{\em Class. Quant.
  Grav.} {\bf 24} (2007)  4169--4190},
\href{http://arxiv.org/abs/0705.4214}{{\tt arXiv:0705.4214 [hep-th]}}.

\bibitem{Sen:2005iz}
A.~Sen, ``{Entropy function for heterotic black holes},''
  \href{http://dx.doi.org/10.1088/1126-6708/2006/03/008}{{\em JHEP} {\bf 03}
  (2006)  008},
\href{http://arxiv.org/abs/hep-th/0508042}{{\tt arXiv:hep-th/0508042
  [hep-th]}}.

\bibitem{Sahoo:2006vz}
B.~Sahoo and A.~Sen, ``{BTZ black hole with Chern-Simons and higher derivative
  terms},'' \href{http://dx.doi.org/10.1088/1126-6708/2006/07/008}{{\em JHEP}
  {\bf 07} (2006)  008},
\href{http://arxiv.org/abs/hep-th/0601228}{{\tt arXiv:hep-th/0601228
  [hep-th]}}.

\bibitem{Sahoo:2006rp}
B.~Sahoo and A.~Sen, ``{Higher derivative corrections to non-supersymmetric
  extremal black holes in N=2 supergravity},''
  \href{http://dx.doi.org/10.1088/1126-6708/2006/09/029}{{\em JHEP} {\bf 09}
  (2006)  029},
\href{http://arxiv.org/abs/hep-th/0603149}{{\tt arXiv:hep-th/0603149
  [hep-th]}}.

\bibitem{Alishahiha:2006ke}
M.~Alishahiha and H.~Ebrahim, ``{Non-supersymmetric attractors and entropy
  function},'' \href{http://dx.doi.org/10.1088/1126-6708/2006/03/003}{{\em
  JHEP} {\bf 03} (2006)  003},
\href{http://arxiv.org/abs/hep-th/0601016}{{\tt arXiv:hep-th/0601016
  [hep-th]}}.

\bibitem{Chandrasekhar:2006kx}
B.~Chandrasekhar, S.~Parvizi, A.~Tavanfar, and H.~Yavartanoo,
  ``{Non-supersymmetric attractors in R**2 gravities},''
  \href{http://dx.doi.org/10.1088/1126-6708/2006/08/004}{{\em JHEP} {\bf 08}
  (2006)  004},
\href{http://arxiv.org/abs/hep-th/0602022}{{\tt arXiv:hep-th/0602022
  [hep-th]}}.

\bibitem{Astefanesei:2006dd}
D.~Astefanesei, K.~Goldstein, R.~P. Jena, A.~Sen, and S.~P. Trivedi,
  ``{Rotating attractors},''
  \href{http://dx.doi.org/10.1088/1126-6708/2006/10/058}{{\em JHEP} {\bf 10}
  (2006)  058},
\href{http://arxiv.org/abs/hep-th/0606244}{{\tt arXiv:hep-th/0606244
  [hep-th]}}.

\bibitem{Kallosh:2006bt}
R.~Kallosh, N.~Sivanandam, and M.~Soroush, ``{The Non-BPS black hole attractor
  equation},'' \href{http://dx.doi.org/10.1088/1126-6708/2006/03/060}{{\em
  JHEP} {\bf 03} (2006)  060},
\href{http://arxiv.org/abs/hep-th/0602005}{{\tt arXiv:hep-th/0602005
  [hep-th]}}.

\bibitem{Majhi:2015pra}
B.~R. Majhi, ``{Entropy function from the gravitational surface action for an
  extremal near horizon black hole},''
  \href{http://dx.doi.org/10.1140/epjc/s10052-015-3744-7}{{\em Eur. Phys. J.}
  {\bf C75} (2015) no.~11, 521},
\href{http://arxiv.org/abs/1503.08973}{{\tt arXiv:1503.08973 [gr-qc]}}.

\bibitem{Gouteraux:2014hca}
B.~Gouteraux, ``{Charge transport in holography with momentum dissipation},''
  \href{http://dx.doi.org/10.1007/JHEP04(2014)181}{{\em JHEP} {\bf 04} (2014)
  181},
\href{http://arxiv.org/abs/1401.5436}{{\tt arXiv:1401.5436 [hep-th]}}.

\bibitem{Donos:2013eha}
A.~Donos and J.~P. Gauntlett, ``{Holographic Q-lattices},''
  \href{http://dx.doi.org/10.1007/JHEP04(2014)040}{{\em JHEP} {\bf 04} (2014)
  040},
\href{http://arxiv.org/abs/1311.3292}{{\tt arXiv:1311.3292 [hep-th]}}.

\bibitem{Charmousis:2009xr}
C.~Charmousis, B.~Gouteraux, and J.~Soda, ``{Einstein-Maxwell-Dilaton theories
  with a Liouville potential},''
  \href{http://dx.doi.org/10.1103/PhysRevD.80.024028}{{\em Phys. Rev.} {\bf
  D80} (2009)  024028},
\href{http://arxiv.org/abs/0905.3337}{{\tt arXiv:0905.3337 [gr-qc]}}.

\bibitem{Charmousis:2010zz}
C.~Charmousis, B.~Gouteraux, B.~S. Kim, E.~Kiritsis, and R.~Meyer, ``{Effective
  Holographic Theories for low-temperature condensed matter systems},''
  \href{http://dx.doi.org/10.1007/JHEP11(2010)151}{{\em JHEP} {\bf 11} (2010)
  151},
\href{http://arxiv.org/abs/1005.4690}{{\tt arXiv:1005.4690 [hep-th]}}.

\bibitem{Moroz:2009kv}
S.~Moroz, ``{Below the Breitenlohner-Freedman bound in the nonrelativistic
  AdS/CFT correspondence},''
  \href{http://dx.doi.org/10.1103/PhysRevD.81.066002}{{\em Phys. Rev.} {\bf
  D81} (2010)  066002},
\href{http://arxiv.org/abs/0911.4060}{{\tt arXiv:0911.4060 [hep-th]}}.

\bibitem{Cadoni:2009xm}
M.~Cadoni, G.~D'Appollonio, and P.~Pani, ``{Phase transitions between
  Reissner-Nordstrom and dilatonic black holes in 4D AdS spacetime},''
  \href{http://dx.doi.org/10.1007/JHEP03(2010)100}{{\em JHEP} {\bf 03} (2010)
  100},
\href{http://arxiv.org/abs/0912.3520}{{\tt arXiv:0912.3520 [hep-th]}}.

\bibitem{Grozdanov:2015djs}
S.~Grozdanov, A.~Lucas, and K.~Schalm, ``{Incoherent thermal transport from
  dirty black holes},''
  \href{http://dx.doi.org/10.1103/PhysRevD.93.061901}{{\em Phys. Rev.} {\bf
  D93} (2016) no.~6, 061901},
\href{http://arxiv.org/abs/1511.05970}{{\tt arXiv:1511.05970 [hep-th]}}.

\bibitem{Gibbons:1987ps}
G.~W. Gibbons and K.-i. Maeda, ``{Black Holes and Membranes in Higher
  Dimensional Theories with Dilaton Fields},''
\href{http://dx.doi.org/10.1016/0550-3213(88)90006-5}{{\em Nucl. Phys.} {\bf
  B298} (1988)  741}.

\bibitem{Garfinkle:1990qj}
D.~Garfinkle, G.~T. Horowitz, and A.~Strominger, ``{Charged black holes in
  string theory},'' \href{http://dx.doi.org/10.1103/PhysRevD.43.3140,
  10.1103/PhysRevD.45.3888}{{\em Phys. Rev.} {\bf D43} (1991)  3140}.
[Erratum: Phys. Rev.D45,3888(1992)].

\bibitem{Goulart:2016cuv}
P.~Goulart, ``{Dyonic black holes and dilaton charge in string theory},''
\href{http://arxiv.org/abs/1611.03093}{{\tt arXiv:1611.03093 [hep-th]}}.

\bibitem{Goulart:2016nkv}
P.~Goulart, ``{Massless black holes and charged wormholes in string theory},''
\href{http://arxiv.org/abs/1611.03164}{{\tt arXiv:1611.03164 [hep-th]}}.

\bibitem{Gauntlett2010}
J.~P. Gauntlett, J.~Sonner, and T.~Wiseman, ``{Quantum criticality and
  holographic superconductors in M-theory},''
  \href{http://dx.doi.org/10.1007/JHEP02(2010)060}{{\em Journal of High Energy
  Physics} {\bf 2010} (2010)  60},
\href{http://arxiv.org/abs/0912.0512}{{\tt arXiv:0912.0512 [hep-th]}}.

\bibitem{Donos2016}
A.~Donos and J.~P. Gauntlett, ``{Minimally packed phases in holography},''
  \href{http://dx.doi.org/10.1007/JHEP03(2016)148}{{\em Journal of High Energy
  Physics} {\bf 2016} (2016)  148},
\href{http://arxiv.org/abs/1512.06861}{{\tt arXiv:1512.06861 [hep-th]}}.

\end{thebibliography}
\end{document}